\documentclass[a4paper,preprint, onecolumn, fleqn]{cas-sc}

\usepackage{lineno,hyperref}

\usepackage[numbers]{natbib}
\usepackage{color}
\usepackage{physics}
\usepackage{url}

\newcommand{\no}[1]{\textcolor{red}{\sout{#1}}}

\def\tsc#1{\csdef{#1}{\textsc{\lowercase{#1}}\xspace}}
\tsc{WGM}
\tsc{QE}
\tsc{EP}
\tsc{PMS}
\tsc{BEC}
\tsc{DE}

\begin{document}
\let\WriteBookmarks\relax
\def\floatpagepagefraction{1}
\def\textpagefraction{.001}
\shorttitle{Deep learning for vibrational wavefunctions}
\shortauthors{Domingo and Borondo}

\title [mode = title]{Deep learning methods for the computation of vibrational wavefunctions}

\author[1,2]{L Domingo}[orcid=0000-0003-3535-3538]
\cormark[1]
\ead{laia.domingo@icmat.es}

\author[1,2]{F Borondo}[orcid=0000-0003-3094-8911]
\cormark[2]
\ead{f.borondo@uam.es}

\credit{Conceptualization of this study, Methodology, Software}

\address[1]{Instituto de Ciencias Matem\'aticas (ICMAT), 
                       Cantoblanco--28049 Madrid, Spain}

\address[2]{Departamento de Qu\'\i mica, Universidad Aut\'onoma de Madrid, 
                        Cantoblanco--28049 Madrid, Spain}

\cortext[cor1]{Corresponding author}
\cortext[cor2]{Principal corresponding author}

\begin{graphicalabstract}
 A deep neural network with a custom loss function is trained to efficiently generate ground and 
 excited wavefunctions for different molecular potentials of interest. \\
 \begin{center}
  \includegraphics[width=0.85\columnwidth]{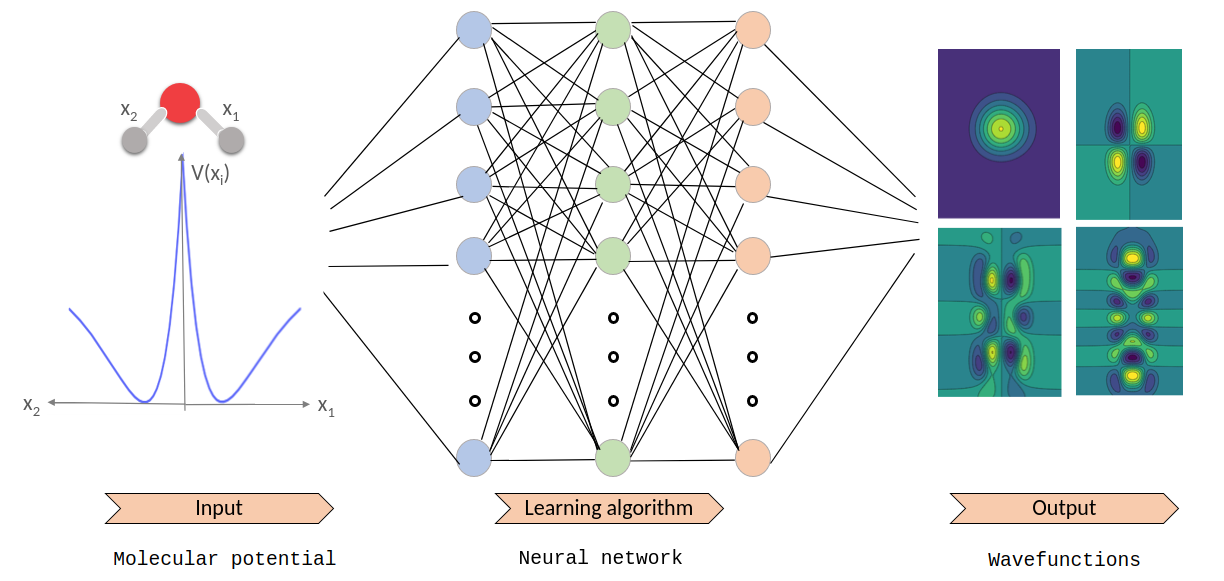}
 \end{center}
\end{graphicalabstract}

\begin{highlights}
\item Use of deep neural networks to solve the time independent Schr\"{o}dinger equation, 
thus obtaining good approximations to the corresponding eigenenergies and eigenfunctions.
\item Modifications of the neural network to make it suitable for obtaining highly excited eigenfunctions, 
where no obvious nodal patterns exist, and quantum numbers are not well defined.
\item Application to typical molecular potentials.
\end{highlights}


\begin{abstract}
In this paper we design and use two Deep Learning models to generate the ground and excited wavefunctions of 
different Hamiltonians suitable for the study the vibrations of molecular systems.
The generated neural networks are trained with Hamiltonians that have analytical solutions, 
and ask the network to generalize these solutions to more complex Hamiltonian functions. 
This approach allows to reproduce the excited vibrational wavefunctions of different molecular potentials. 
All methodologies used here are data-driven, therefore they do not assume any information about the 
underlying physical model of the system. 
This makes this approach versatile, and can be used in the study of multiple systems in quantum chemistry. 
\end{abstract}

\maketitle

\section{Introduction}
  \label{sec:intro}
The accurate computation of the eigenstates of a dynamical system is a central problem in physics and 
computational chemistry.
For example, being able to efficiently solve the Schr\"{o}dinger equation is crucial to determine molecular 
structural properties and molecular dynamics in any quantum-mechanical scenario \cite{Lanyon}. 
When the size and/or complexity of the system increases, finding such solutions becomes challenging. 
The usual standard methods \cite{CompChem} are based on the variational principle, which implies that in order 
to get an approximation to the $N$-th eigenstate, the $N-1$ lower-lying ones should also be calculated. 
This makes the task particularly demanding when one is interested in excited states.  
The existence of a large density of states and/or classically chaotic dynamics, as it happens in the molecular
rovibrational case \cite{Revuelta1}, as opposed to the molecular Born-Oppenheimer electronic one \cite{CompChem}, 
also significantly contributes to the computational burden \cite{Revuelta2}.
For this reason, many methods have have been proposed in the literature to either approximate the underlying 
interactions of the system \citep{Vrubel,Ipatov} or to numerically approximate the exact
Schr\"{o}dinger equation \cite{Revuelta3}. 
These methods are based on finding an appropriate description of the system dynamics.

Another option is the use data-based approaches, which do not rely on information about the underlying 
physical model of the system, but use only data obtained from observations of the system. 
An increasingly popular family of these data-based methods is machine learning (ML).
Indeed, exciting recent work has been devoted to the use of ML techniques to study partial differential
equations \citep{E,Ruthotto,SIAMnews}, such as the Schr\"{o}dinger equation \cite{Delin}, 
which are at the core of practically all branches of science.

ML is an emerging  mathematical and computer science field of study which aims to give computers
the ability to learn from examples and experience, without being explicitly programmed to solve 
the particular task under study. 
ML is nowadays present in many areas of technology, and any user of today's technology heavily 
benefits from its applications, often without even being aware it.
Among the most popular and widespread applications of ML
facial recognition technology \citep{Ruthotto,Application4}, which allows social media platforms to help 
users tag and share photos of friends, 
effective web search \cite{Application1}, which eases the acquirement of information, or 
self-driving cars \cite{Application2}, which will soon be available to customers, are worth mentioning.

One of the most popular and widespread ML method is Artificial Neural Networks (ANN) \cite{ANN} 
[or just Neural Networks (NN)], and in particular Deep Learning (DL) \citep{DeepL1,DeepL2}. 
ANN is a widespread method used for generalization problems, especially for the nonlinear function
approximation \cite{DLnonlinear}. 
An ANN is an information processing paradigm inspired by the way that biological nervous
systems, particularly the human brain,  process information. 
These NNs are composed of a large number of interconnected ``neurons'' or nodes, which work 
coordinately to solve specific problems. 
ANN are considered to be deep (DNN) when formed by a large number of neuron layers. 
In this way, DL consists of using deep ANN as the learning algorithm.

In the field of computational chemistry, ML has been extensively used \cite{SpecialIssue}
to solve the electronic \citep{Noe}, and,
to a lesser extent, the vibrational Schr\"{o}dinger equations \citep{RigoMLTST,Schnet, Schnet2}, 
as well as to compute Born-Oppenheimer potential energy surfaces \citep{Bowman,Schran},
to design new materials \citep{RigoML,Pollice},
and to elucidate the form of a Hamiltonian from its eigenfunctions (inverse problem) \cite{InverseProblem, PINN}. 
In the case of the vibrational Schr\"{o}dinger equation, DL methods have proven to give remarkable results in 
predicting the ground energy of multiple Hamiltonians \cite{DLSchrodinger, MLSchrodinger, CompetingLosses},
but there are no relevant applications to the excited case, which is much more interesting since controlled by
anharmonicities and mode couplings, and often give rise to the so-called scarred functions \cite{Revuelta3}.

In this paper, instead of predicting the mean energy of an eigenstate, we will obtain the full wavefunction for such state, 
which provides full information about the system state. 
Moreover, instead of only focusing on the ground state of the Hamiltonian, we will also obtain high lying states, 
which correspond to more complex wavefunction topologies. 
Two different scenarios will be studied. 
In the first one, we use polynomial potentials and their associated eigenfunctions to train a neural network. 
Then, the network is asked to generalize to non-polynomial potentials. 
In the second scenario, we start by using molecular potentials with analytical eigenstates to train the network,
and then test the network with more complex perturbed potentials, which have no analytical solution. 

The organization of this paper is as follows. 
In Sect.~\ref{sec:modelI} we introduce the neural network model and training details for random polynomial 
potentials. 
Similarly, we present in Sect.~\ref{sec:modelII} the neural network model and training details for the molecular 
potentials under study. 
The results for both cases are presented and discussed in Sect.~\ref{sec:results}. 
Finally, Sect.~\ref{sec:summary} ends the paper by summarizing the main conclusions of the present work.

\section{Models and Methods}
 \label{sec:methods}
 In this section we present and discuss the potentials and method used for the two scenarios studied in this paper,
 i.e.,~random polynomial potentials (Model I) and coupled Morse potentials applied to the H$_2$O molecule (Model II).
\subsection{Model I: Random polynomial potentials}
 \label{sec:modelI}

The general goal of this work is to train a NN to generate the ground and excited eigenfunctions of different molecular 
vibrational Hamiltonians. 
For this purpose, the NN is trained with Hamiltonians belonging to the same family of functions. 
Then, we expect our NN to be able to generalize and reproduce the wave functions of Hamiltonians described by more 
general expressions.
In this work, we consider both one-dimensional (1D) and a two-dimensional (2D) potentials.
In order to design our model, we need to specify both the training data and the learning algorithm. 

In the first place (model I), we will consider that the training data is a set of Hamiltonians with random polynomial 
potentials of (up to) degree four. That is, for the 1D case
\begin{equation}
H(x) = \frac{p^2}{2m} + V(x), \qquad \mbox{with} \quad 
          V(x) = 
          \sum_{i\leq4} \alpha_i x^i,
\label{eq:1}
\end{equation}
and for 2D
\begin{equation}
H(x,y) = \frac{p_x^2 + p_y^2}{2m} + V(x,y), \qquad \mbox{with} \quad V(x,y) = \sum_{i+j\leq4} \alpha_{ij} x^i y^j .
\label{eq:2}
\end{equation}
Each Hamiltonian has an associated set of eigenfunctions, which are the solution of the corresponding time-independent 
Schr\"{o}dinger equation
\begin{equation}
H(\vec{r}) \ \psi(\vec{r}) = E \ \psi(\vec{r}), \quad E \in \mathbb{R},
\label{eq:3}
\end{equation}
which in our case will be obtained with the variational method described below.

Since the kinetic energy operator is the same for all Hamiltonians, we only have to provide the potential function  
to the NN. 
This allows to pass an easy representation of the Hamiltonian to the NN. 
Therefore, the training data consists of a set of pairs $\{V_i, \psi_i\}_i$, where $V_i$ is the $i$-th training potential and $\psi_i$ 
the associated computed wave function.
Both, potentials and wave functions, are represented in a grid on a rectangular (or linear) domain,
so that $V_i$ is a matrix containing the values of $V_i (x,y)$ [or $V_i (x)$ for 1D] with $x,y$ belonging to a rectangular lattice 
(or a closed interval for 1D). 
Similarly, $\psi_i$ is a matrix containing the values of $\psi(x,y)$ in such lattice (or closed interval for 1D).

\begin{figure}
\centering
  \includegraphics[width=0.5\columnwidth]{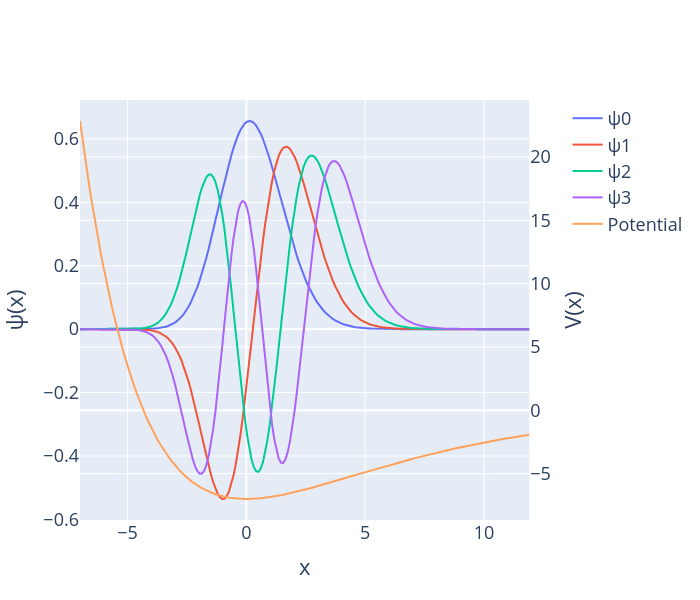}
 \caption{Example of a Morse potential with $D_e=7$, $a=0.16$ and $x_e = 0$, and the corresponding eigenfunctions 
 for the energy levels $n=0,1,2,3$.
}
 \label{fig:1}
\end{figure}

Once the NN has been trained to reproduce the different wave functions for a polynomial potential for a particular energy vibrational 
state (either the ground or excited state), we ask the network to reproduce the wave functions for another more general, 
non-polynomial potential. 
In this work, we chose to test the network against Morse potentials, which are sufficiently different from the training potentials, 
and they also adequately represent the potential interaction of a diatomic molecule. We write the Morse potential as
\begin{equation}
V(x) = D_e \left[e^{-2a(x-x_e)} - 2e^{-a(x-x_e)}\right],
\label{eq:4}
\end{equation}   
where $x$ is the distance between atoms, $x_{e}$ is the corresponding equilibrium bond distance, 
$D_{e}$ is the well depth (defined relative to the dissociated atoms), and $a$ is a parameter controlling the ``width'' of the potential 
(the smaller $a$ is, the deeper the well). 
This potential approaches zero at $x \rightarrow \infty$ and equals $-D_{e}$ at its minimum at $x=x_e$. 
The Morse potential is the combination of a short-range repulsion term (the former) and a long-range attractive term (the latter). 
The Hamiltonian associated to the Morse potential has analytical solution for the eigenenergies $\{E_n\}$ and eigenfunctions $\{\phi_n(x)\}$,
$n$ being the corresponding quantum number, which are given by
\begin{equation}
E_n = -\frac{a^2 \hbar^2}{2m} \left(\lambda - n - \frac{1}{2}\right)^2, \quad n=0,1,2, \cdots, \left[\lambda - \frac{1}{2}\right],
  \label{eq:5}
\end{equation}
and
\begin{equation}
\phi_n(z) = N_n z^{\lambda - n - 1/2} e^{-1/2z} L_n^{(2\lambda - 2n -1)}(z) ,
  \label{eq:6}
\end{equation}
respectively, where:
\begin{equation}
\lambda = \displaystyle\frac{\sqrt{2mD_e}}{a \hbar}, \qquad 
z = 2\lambda e^{-a(x-x_e)}, \qquad  \mbox{and} \quad
N_n = \Big(\displaystyle\frac{n!(2\lambda -2n -1)}{\Gamma(2\lambda -n)} \Big)^{1/2},\vspace{0.07in}\\
  \label{eq:7}
\end{equation}
and $L_n^{(\alpha)}$ is a generalized Laguerre polynomial. 
Figure \ref{fig:1} shows an example of a Morse potential, together with the eigenfunctions for the first four lowest energy levels. 

\subsubsection{Data Generation}
  \label{sec:data_poly}
\begin{figure}
\centering
  \includegraphics[width=0.85\columnwidth]{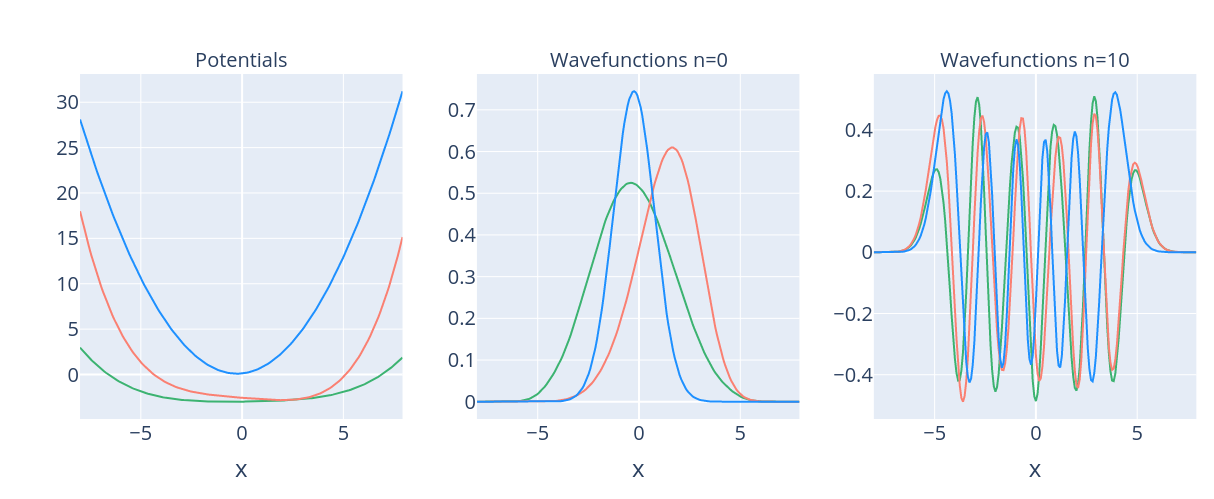}
 \caption{Three examples of random polynomial potentials (left) and their associated eigenfunctions for the ground
 state (middle) and the 11-th excited state (right) obtained with the variational method described in the text.}
 \label{fig:2}
\end{figure}

As explained in section \ref{sec:modelI}, the NN model is trained using polynomial potentials up to degree four. 
To ensure that the eigenstates have discrete energies, and thus are physical bound states (as opposed to continuum states),
we impose some restriction properties on the coefficients.
In our case, we will make sure that the even terms ($x^2$ and $x^4$) dominate over the odd terms ($x$ and $x^3$). 
Also, we allow the potential to be negative and non-centered by including negative values 
for $\alpha_0$ and $\alpha_1$ [see Eq.~(\ref{eq:1})]. 
Finally, we use small values of the coefficients so that the potential does not achieve very high values, 
which can lead to numerical instability. 
The values of $\{\alpha_i\}$ (for 1D potentials) and $\{\alpha_{ij}\}$ (for 2D potentials), 
chosen according to the previous conditions, are shown in Table \ref{table:1}. 
\begin{table}
\begin{tabular}{ccc}
\hline 
$\alpha_i$ & min & max\\
\hline
$\alpha_0$ & -4.5 & 1.5\\
$\alpha_1$ & -0.65 & 0.65\\
$\alpha_2$ & 0.2 & 1.0\\
$\alpha_3$ & -0.01 & 0.01\\
$\alpha_4$ & 0 & 0.1\\
\hline
\end{tabular}
\qquad
\begin{tabular}{ccccccccc}
\hline
$\alpha_{ij}$ & min & max    & \qquad$\alpha_{ij}$   & min    & max & \qquad$\alpha_{ij}$  & min   & max\\
\hline 
$\alpha_{00}$ & -3 & 0.1      & \qquad$\alpha_{02}$ & 0.2    & 1.0  & \qquad$\alpha_{04}$ & 0      & 0.2 \\
$\alpha_{10}$ & -0.2 & 0.1    & \qquad$\alpha_{21}$ & -0.02 & 0.02 & \qquad$\alpha_{13}$ & -0.01 & 0.01 \\
$\alpha_{01}$ & -0.2 & 0.1    & \qquad$\alpha_{12}$ & -0.01 & 0.01 & \qquad$\alpha_{22}$ & 0      & 0.04 \\
$\alpha_{11}$ & -0.02 & 0.02 & \qquad$\alpha_{03}$ & -0.01 & 0.01 & \qquad$\alpha_{31}$ & -0.01 & 0.01 \\
$\alpha_{20}$ & -0.05 & 0.05 & \qquad$\alpha_{30}$ & -0.01 & 0.01 & \qquad$\alpha_{40}$ & 0      & 0.02 \\
\hline

\end{tabular}
\caption{Lower and upper bounds for the coefficients of the polynomial potentials in one dimension 
of Eq.~(\ref{eq:1}) (left) and in two dimensions of Eq.~(\ref{eq:2}) (right). }
\label{table:1}
\end{table}
Figure \ref{fig:2} shows some examples of random polynomial potentials and their associated wavefunctions for 
two values of the vibrational number $n=0$ and $n=10$.

Apart from the polynomial potentials, the training set also contains the eigenfunctions associated to such Hamiltonians. 
These random Hamiltonians do not usually have an analytical solution, hence a numerical solver needs to be used. 
In this work, we use the variational method using harmonic oscillator eigenfunctions $\phi_n(x)$ as a basis set
to generate the eigenfunctions of an arbitrary Hamiltonian $H$. 
That is, since $\{\phi_n(x)\}$ form a complete basis set for the Hilbert space $\mathcal{H}$  we can write any 
wavefunction $\psi(x)\in\mathcal{H}$ as a linear combination of the harmonic oscillator eigenfunctions
\begin{equation}
\psi(x) = \sum_{n=0}^\infty a_n \phi_n(x), \quad a_n \in \mathbb{R} \ \forall n
\label{eq:8}
\end{equation}
Therefore, the problem reduces to find the values of $\{a_n\}$ which minimize the expected energy 
\begin{equation}
\expval{H} = \expval{H}{\psi} = \int_{-\infty}^\infty \left[\sum_{n=0}^\infty a_n \phi_n(x)\right] H 
     \left[\sum_{m=0}^\infty a_m \phi_m(x)\right] \, dx,
\label{eq:9}
\end{equation}
where it is assumed that the eigenfunction $\psi$ is normalized in the standard way, i.e.,~$\langle\psi|\psi\rangle=1$.
For full details of the variational method, and its applications to our 2D potentials, see appendices \ref{sec:appendixA} 
and \ref{sec:appendixB}.
\subsubsection{Neural network}
  \label{sec:NN_poly}

In this work, we use NNs as the learning algorithms for both the 1D and 2D problems described above. 
Here we describe the architecture and learning process in both cases.

\begin{figure}
\centering
  \includegraphics[width=0.85\columnwidth]{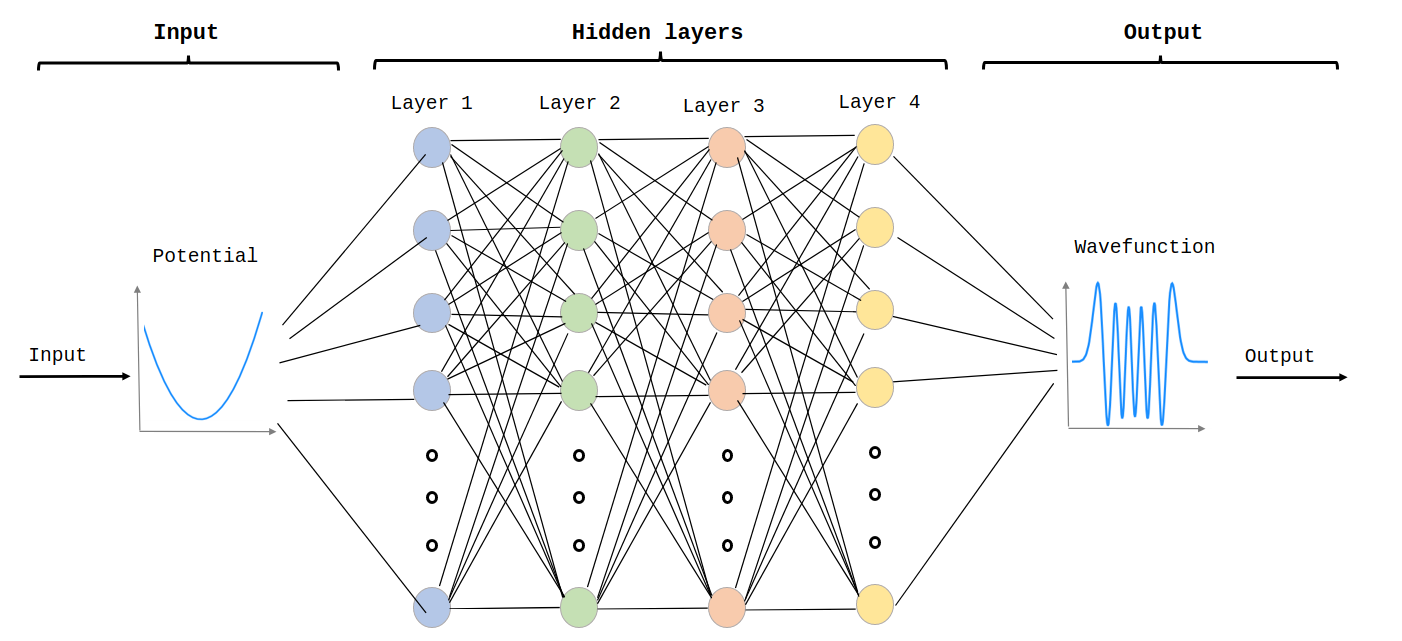}\\ \vspace*{0.5cm}
  \includegraphics[width=0.85\columnwidth]{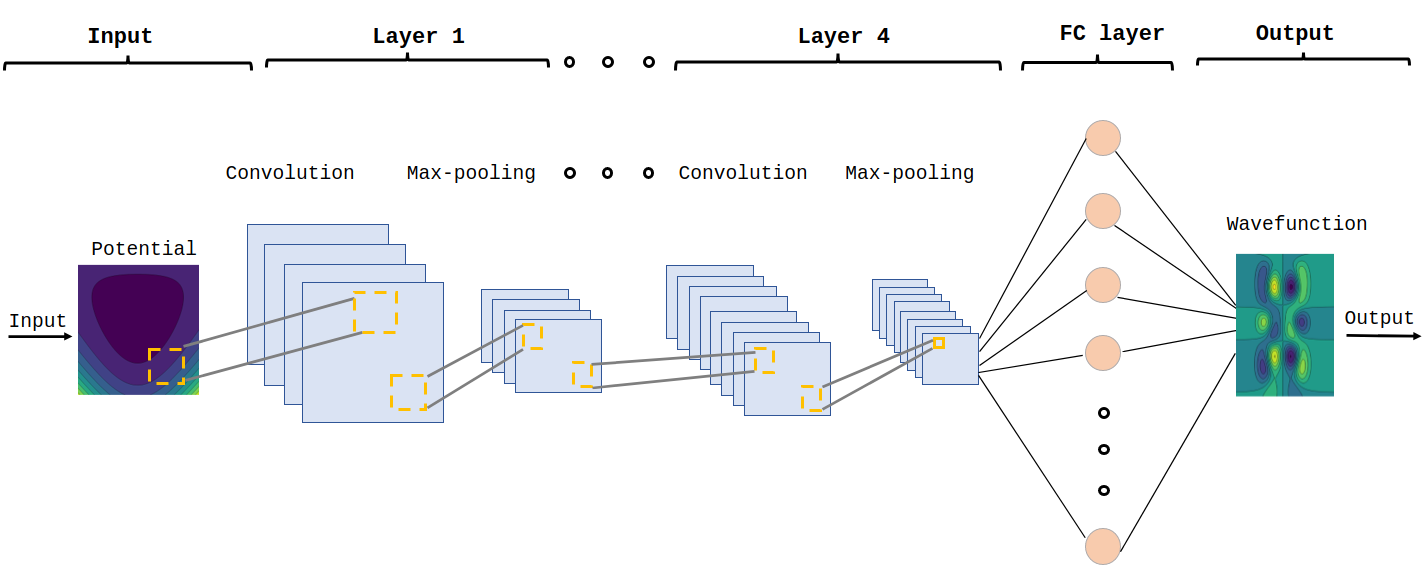}
 \caption{Schematic plots of the two types of neural networks considered in this paper:
   Fully connected neural network (FCNN) used  for the 1D potentials (upper panel),
   and convolutional neural network (CNN) used for 2D potentials (bottom panel). 
   The FCNN is composed by an input layer, four fully-connected layers and an output layer. 
   The input is a 1D array containing the potential $V(x)$, and the output is its associated wavefunction for a particular quantum number $n$. 
   The CNN is composed by the input layer, four convolutional and max-pooling layers, a fully-connected layer and an output layer. 
   The input is a 2D array containing the potential $V(x,y)$ and the output is its associated wavefunction for a pair of quantum 
   numbers ($n_x,n_y$).  }
 \label{fig:NN}
\end{figure}

For 1D potentials, we use a fully connected neural network (FCNN). 
The input is an array of 200 points containing the values of the potential in the spatial domain $x\in[-8,8]$ for the 
fundamental state and $x\in[-20,20]$ for the excited states. 
Our NN consists of four fully connected layers with 256, 256, 128 and 128 neurons respectively, and RELU activation 
functions \cite{Relu}. 
Every FC layer is followed by a dropout layer with parameter 0.2. 
These dropout layers help prevent overfitting and thus help the network to generalize to unseen potentials. 
The output layer is a linear layer with 200 neurons, which predicts the wave function for the given potential. 
A schematic plot of our FCNN is displayed in Fig.~\ref{fig:NN}
This network is trained using an Adam optimizer \cite{Adam} with a learning rate of 0.0005 for 1000 iterations. 
Moreover, we used early stopping to further prevent overfitting the network. 
The network was trained with 5000 samples with a batch size of 64.
%

For 2D potentials, we use a convolutional neural network (CNN).
This model is used to extract a lower-dimensional embedding of the original data which is then used to make the predictions. 
CNNs are known to be very useful to extract local patterns from the input data and to extract valuable features for the learning task. 
The input in this case is a 2D array of size 100$\times$100 in the spatial domain $(x,y)\in[-10,10] \times [-10,10]$. 
In this case, we use four convolutional layers with 64, 64, 32 and 32 filters respectively.  
The kernel size is 3 for all four layers, and the stride is (2,2).  
All activation functions are also RELU here. 
After each convolutional layer, we add a max-pooling layer to reduce the dimensionality of the embedding. 
We use a pooling size of (2,2) and a stride of (1,1). 
Moreover, after each max-pooling layer we add again a dropout layer with parameter 0.2 to avoid overfitting the network. 
Then, we add 2 fully connected layers with 128 neurons each. 
The output layer is a linear layer of the same size as the input. 
The training is performed in the same way as in the 1D case. 
A schematic representation of the CNN is displayed in Figure \ref{fig:NN}. 

\subsection{Model II: Morse potentials}
  \label{sec:modelII}
The goal of the second part of the work is to study the performance of a NN trained with the solutions of an
analytically solvable Hamiltonian $H_0$ in reproducing the wave functions of the corresponding perturbed 
non-separable Hamiltonian $H$. For this purpose, let $H$ be the Hamiltonian whose eigenfunctions we want to find, 
and suppose that it can be written \textit{\`a la Kolmogorov-Arnold-Moser} \cite{Llave} as
\begin{equation}
H = H_0 + H_1,
\label{eq:10}
\end{equation}
where $H_0$ is a Hamiltonian whose eigenstates are analytically known. 
If the ``perturbation'' $H_1$ is small, then $H_0$ is a good approximation of $H$, and it can be expected that a NN
can generalize the wavefunctions for $H_0$ to those for $H$.
However, this is not obvious \textit{a priori} since resonance between modes in the excited states can change significantly the
topology of these wavefunctions.

\subsubsection{Coupled Morse potentials}
  \label{sec:coupled_morse}

We consider the following kinetically coupled 2D Morse oscillator, which has been extensively studied in the past
as a model for the stretching vibrations of the H$_2$O molecule \citep{Jaffe,H2O2,H2O,H20_NN}
\begin{equation}
H(x_1, x_2, p_1, p_2) = \frac{1}{2}(G_{11}p_1^2 + G_{22}p_2^2) + G_{12}p_1p_2 + U_M(x_1) + U_M(x_2),
\label{eq:11}
\end{equation}
being the $G$-matrix elements equal to
\begin{equation}
G_{11} = G_{22} = \frac{m_H + m_O}{m_Hm_O}, \quad G_{12} = \frac{\cos\alpha}{m_O},
\end{equation} 
where $m_H = 1.00784$ amu  and $m_O = 15.999$ amu are the H and O atomic masses, respectively, 
and $\alpha$ is the bending angle which in this model is held frozen at 104.5$^\circ$.
Functions $U_M(x_{1,2})$ are 1D Morse potentials in the stretching H--O coordinates, $x_1$ and $x_2$, characterized
by parameters $a$ and $D_e$, and $p_1, p_2$ are the corresponding conjugate momenta.
For the H$_2$O molecule $G_{12} \approx 0.01559$, and therefore $H_0$ is a good approximation of $H$, so that
Hamiltonian (\ref{eq:11}) can be written in the form (\ref{eq:10}) by making
\begin{equation}
  H_0 = \frac{1}{2}(G_{11}p_1^2 + G_{22}p_2^2) + U_M(x_1) + U_M(x_2) 
    \qquad \mbox {and} \quad H_1 = G_{12}p_1p_2.
  \label{eq:13}
\end{equation}

As explained in the previous section, the way in which we perform our study consists in providing only the potential function to the NN, 
instead of giving the whole Hamiltonian. 
This fact allows to have an easy representation of the Hamiltonian as a grid containing the values of the potential energy 
on a rectangular spatial domain. 
This representation was possible in the cases presented before in Sect.~\ref{sec:NN_poly} because all the Hamiltonians 
had the same kinetic energy. 
Nonetheless, our coupled Morse Hamiltonian is different since contains a coupling term in the momentum coordinates 
and not in the spatial coordinates. 
Therefore, to be able to provide only the potential energy to the NN, we have to make a change of coordinates in the
Hamiltonian in such a way that the coupling appears only in the spatial part, i.e.,~the potential.
To this end, we can rewrite the Hamitonian in generalized coordinates so that the coupling appears in the spatial coordinates 
instead of in the momentum coordinates.

In order to do so, we apply a canonical transformation to the Hamiltonian $H(x_1,x_2,p_1,p_2)$ 
obtaining the Hamiltonian $H'(x_1', x'_2, p_1', p_2')$. 
We use a generating function of the form $F_2(x_1, x_2, p_1', p_2')$ \cite{Goldstein}, so that
\begin{equation}
\begin{array}{l}
p_i = \displaystyle\frac{\partial F_2}{\partial x_i}, \qquad i=1,2\vspace{0.07in}\\
x_i' = \displaystyle\frac{\partial F_2}{\partial p_i'}, \qquad i=1,2\vspace{0.07in}\\
H' = H + \displaystyle\frac{\partial F_2}{\partial t}.
\end{array} 
\end{equation}
In particular, by choosing $F_2(x_1,x_2, p_1', p_2')$ of the form
\begin{equation}
F_2(x_1,x_2, p_1', p_2') = f_1(x_1,x_2)\,p_1' + f_2(x_1,x_2)\,p_2'
\end{equation}
then, the generalized coordinates fulfill the following equations
\begin{equation}
\begin{array}{ll}
p_1 = \displaystyle\frac{\partial F_2}{\partial x_1} = \frac{\partial f_1}{\partial x_1}p_1' 
  +\frac{\partial f_2}{\partial x_1}p_2'\vspace{0.07in} \\
p_2 = \displaystyle\frac{\partial F_2}{\partial x_2} = \frac{\partial f_1}{\partial x_2}p_1' 
  +\frac{\partial f_2}{\partial x_2}p_2'\vspace{0.07in}\\
x_1' = \displaystyle\frac{\partial F_2}{\partial p_1'} = f_1(x_1,x_2)\vspace{0.07in}\\
x_2' = \displaystyle\frac{\partial F_2}{\partial p_2'} = f_2(x_1,x_2).
\end{array} 
\label{eq:16}
\end{equation}
In this way, the kinetic energy can be written in matrix form as
\begin{equation}
T(p_1,p_2) = \frac{1}{2} \begin{pmatrix}
p1 & p2\\
\end{pmatrix} 
\begin{pmatrix}
G_{11} & G_{12}\\
G_{12} & G_{11}
\end{pmatrix}
\begin{pmatrix}
p1\\
p2\\
\end{pmatrix} := \begin{pmatrix}
p1 & p2\\
\end{pmatrix} 
M
\begin{pmatrix}
p1\\
p2\\
\end{pmatrix},
\end{equation}
so that, diagonalizing $M$
\begin{equation}
M = S D  S^T \qquad \mbox{with} \quad
S = \frac{1}{\sqrt{2}}
\begin{pmatrix}
-1 & 1\\
1 & 1\\
\end{pmatrix}, \qquad \mbox{then} \quad
D = 1/2 \begin{pmatrix}
G_{11} - G_{12} & 0\\
0 & G_{11} + G_{12}
\end{pmatrix}.
\end{equation}
Accordingly, defining new coordinates $\vec{p'} = S^T \vec{p}$, the new kinetic energy only consists of by diagonal terms
\begin{equation}
T'(p_1', p_2') = \frac{1}{2} \left(G_{11}-G_{12}\right) \, p_1'^2 + \frac{1}{2} \left(G_{11}+G_{12}\right) \, p_2'^2.
\end{equation}
Now, the transformation defined by Eqs.~(\ref{eq:16}) can be used to compute the new set of coordinates $(x_1', x_2')$
%
\begin{equation}
x_1' = f_1(x_1,x_2) = \frac{1}{\sqrt{2}}(-x_1 + x_2), \qquad x_2' = f_2(x_1,x_2) = \frac{1}{\sqrt{2}}(x_1 + x_2),
\end{equation}
to obtain the new Hamiltonian
\begin{equation}
H'(x_1', x_2', p_1', p_2') = \frac{1}{2} \left(G_{11}-G_{12}\right)\,p_1'^2 + 
                                     \frac{1}{2} \left(G_{11}+G_{12}\right)\,p_2'^2 + 
                                     U_M\left(\frac{1}{\sqrt{2}}\left[x_2' - x_1'\right]\right) + 
                                     U_M\left(\frac{1}{\sqrt{2}}\left[x_2' + x_1'\right]\right) .
\end{equation}
Finally, scaling the coordinates so that both particles have the same mass, we define
\begin{equation}
x = \frac{1}{\sqrt{G_{11} - G_{12}}}x_1', \quad y = \frac{1}{\sqrt{G_{11} + G_{12}}}x_2', \quad 
      p_x = \dot{x}, \quad p_y = \dot{y},
\end{equation}
obtaining the Hamiltonian with the coupling in the spatial coordinates as
\begin{multline}
K(x,y,p_x,p_y) = \frac{1}{2}(p_x^2 + p_y^2) + 
   U_M\left(\frac{1}{\sqrt{2}}\left[\sqrt{G_{11} + G_{12}}y- \sqrt{G_{11} - G_{12}}x\right]\right) + \\
   U_M\left(\frac{1}{\sqrt{2}}\left[\sqrt{G_{11} + G_{12}}y+ \sqrt{G_{11} - G_{12}}x\right]\right).
 \label{eq:23}
\end{multline}
Let us remark, that this equivalent Hamiltonian has no kinetic coupling in the momenta $p_x$ and $p_y$, 
but it has been moved to the potential term, between spatial coordinates $x$ and $y$,
which is more adequate for our computational purposes when using NN, as indicated before. 

To further illustrate the effect of the above transformation, we compute the Taylor expansion of the new Hamiltonian,
thus obtaining
\begin{multline}
K(x,y,p_x,p_y) = \frac{1}{2}(p_x^2 + p_y^2) -2D + D_ea^2(G_{11}-G_{12})x^2 + D_ea^2(G_{11}+G_{12})y^2 - \\ 
 \frac{3}{\sqrt{2}}D_ea^3(G_{11}-G_{12})\sqrt{G_{11} + G_{12}}yx^2 - \frac{1}{\sqrt{2}}Da^3 (G_{11}+G_{12})^{3/2} y^3 + 
 {\cal O}(4)
\label{eq:24}
\end{multline}
where we see reappearing the two Morse parameters $D_e$ and $a$. 
We observe that the expansion contains a coupling term $yx^2$, and an anharmonicity in $y^3$, 
this indicating that up to order 3 this Hamiltonian is identical to that proposed by H\'enon and Heiles to study the stability of some galaxies, 
at the dawn of nonlinear science \cite{HH}.

Now that we have defined the coupled Morse potential in generalized coordinates, we have to define the training data of the NN. 
In this case we use decoupled Morse potentials to train the network, which act as a first order approximation of the coupled Morse potential. 
The training potentials will then be of the form
\begin{equation}
V_0(x,y) = U_M^{D_1,a_1}(x) + U_M^{D_2,a_2}(y),
\label{eq:25}
\end{equation}
where $U_M^{D,a}$ represents a Morse potential with well depth $D$ and width $a$. 
Notice that the Morse parameters of this Hamiltonian are different in each spatial coordinate. 
The Hamiltonian associated to this potential is separable, and the Schr\"{o}dinger equation has analytical solution in terms of 
expressions (\ref{eq:5})-(\ref{eq:7}). 
Thus, there is no need to use a numerical solver to train the NN, which makes the method more convenient. 
In this case the training data are pairs $\{V_i,\psi_i\}$, where $V_i$ is a matrix containing the values of the potential in a rectangular domain, 
and $\psi_i$ is the associated wavefunction corresponding to the vibrational quantum numbers $n_x,n_y$. 
We train different models for different quantum numbers in order to reproduce multiple excited eigenstates of the coupled Morse Hamiltonian. 
Notice that here the excited states are identified by their quantum numbers $n_x$, $n_y$ instead of their energies $E(n_x,n_y)$. 
We made this choice because when the difference between eigenenergies is small, the $n$th energy level of two similar potentials 
can correspond to very different quantum numbers $n_x$, $n_y$, and consequently completely different wave functions. 
This fact would ``confuse'' the NN since two similar inputs would have very different outputs. 
In the next section we give further details about how the training data are generated.

\subsubsection{Data Generation}
  \label{sec:data_coupled}
Neural networks are useful methods for extracting features of complex data. 
However, if the training and test sets are too different, the network may not be able to produce good results. 
For this reason, it is important that the training data resembles as much as possible the test data. 
In this work we are approximating a coupled Morse Hamiltonian with a separable Morse Hamiltonian [see Eq.~(\ref{eq:25})]. 
Two strategies were then used to generate useful training data:

\begin{itemize}
\item \textbf{Selecting the Morse parameters $(D_1,a_1)$, $(D_2,a_2)$:} 
We selected the parameters of the decoupled potential in two different ways:

\begin{itemize}
\item By performing curve fitting and finding the parameters $(D_1,a_1)$, $(D_2,a_2)$ which best approximate the coupled Morse 
potential in Eq.~(\ref{eq:10}).
\item By choosing the parameters $(D_1,a_1)$, $(D_2,a_2)$ which have the same quadratic order Taylor expansion coefficients 
as the coupled Morse potential. 
Considering Eq.~(\ref{eq:24}) the choice should be
\begin{equation}
D_1 = D_2 =D_e  \qquad \mbox{and} \quad 
a_1 = \sqrt{G_{11} - G_{12}} \quad 
a_2 = \sqrt{G_{11} + G_{12}}a\
\end{equation}
\end{itemize}
In order to obtain the training data we first generated samples of the parameters of the coupled Morse potential 
$a\in [0.09, 0.12]$, $D_e \in [1,10]$, and then found the decoupled Morse $(D_1,a_1)$, $(D_2,a_2)$ parameters, 
according to the previous strategies.  

\item \textbf{Selecting the $x$ and $y$ ranges:} Once the Morse parameters have been chosen, we try to improve the resemblance 
with the coupled Morse potential by changing the values of the spatial domains. 
The input of the NN is a grid containing the values of the potential in a certain spatial domain, but such domain is not specifically given. 
Therefore, if we change the range of this domain the network will not notice the difference, as long as the number of points remains constant. 
This fact allows us to stretch the spatial domain so that the decoupled potential is more similar to the coupled potential. 
We performed a grid search to find the domain ranges which best approximate the coupled potential energy. 
The only constraint is that in a given spatial domain, the associated wave function fits into that domain. 
Otherwise, the sample will no be useful for training. 
Recall that this technique could only be used because the learning algorithm, i.e.,~the NN is a data-based approach, 
instead of a model-based approach, which means that uses no information about the underlying physical model of the system. 
\end{itemize}

\subsubsection{Neural network}
  \label{sec:NN_coupled}

The architecture of the NN in this case is the same as the one used with the 2D random polynomial potentials 
in Sect.~\ref{sec:NN_poly}. 
The only difference in the training process is the choice of the loss function \cite{LossFnt}.
When the energy of the system increases, the eigenfunctions of the coupled Morse Hamiltonian show significant differences 
to any of the eigenfunctions of the decoupled Morse Hamiltonian, due to the effect of the different nonlinear resonances existing
in the system \citep{Jaffe,H2O2}. 
For example, the number of nodes of the wavefunction may not be well-defined. 
In this case, training the neural network with only the decoupled wavefunction does not give optimal results. 
For this reason, we add a custom loss function to help training the NN, defined in the following way 
%
\begin{equation}
  S_{loss} = \norm{H\widetilde{\psi} - \widetilde{E}\widetilde{\psi}}^2 + \lambda_\text{norm}\norm{\widetilde{\psi}}^2,
\end{equation}
where $\widetilde{E}$ is the predicted mean energy, which is calculated using the input potential and the predicted 
wavefunction $\widetilde{\psi}$. Therefore the total loss function is
\begin{equation}
\mathcal{L} =  MSE_d + \lambda \bar{S}_\text{loss,c} =  \frac{1}{N_d} \sum_{i=0}^{N_d} \norm{\widetilde{\psi}_i^d - \psi_i^d}^2 
   + \lambda \sum_{i=0}^{N_c} \left(\norm{H\widetilde{\psi}_i^c - \widetilde{E}\widetilde{\psi}_i^c}^2 + \lambda_\text{norm}\norm{\widetilde{\psi}_i^c}\right),
\end{equation}
where $\{\psi_i^d\}_i$ are the wavefunctions of the decoupled Hamiltonian and $\{\psi_i^c\}_i$ are the wavefunctions of the 
coupled Hamiltonian, and $\lambda, \lambda_{norm} \in \mathbb{R}$. 
Parameter $\lambda_{norm}$ is chosen so that the two terms of the Schr\"{o}dinger equation loss have the same order of magnitude. 
In this case, we choose $\lambda_{norm}=10^4$ for all the training process. 
We train the network for 300 iterations. 
During the first 100 iterations we set $\lambda=0$ so that the model learns to reproduce the wavefunctions of multiple 
decoupled Hamiltonians. 
Then, we choose $\lambda$ so that the two loss functions have the same order of magnitude. 
In this case, we set $\lambda=10^5$. 
Notice that since the decoupled Morse potential already gives a fair approximation of the true wavefunction, 
we do not need to put any constraints on the energy of the system. 
The NN converges to the true solution, which follows the Schr\"{o}dinger equation.

\section{Results}
  \label{sec:results}
\subsection{Model I: Polynomial potentials}
  \label{sec:res-poly}

As described in the previous section, in the first part of this work we use random polynomial potentials
and their associated eigenfunctions, to train our NN. 
Afterwards, we test the ability of the network to generalize to non-polynomial potentials, 
in particular to the Morse potential. 

\begin{figure}
\centering
  \includegraphics[width=0.85\columnwidth]{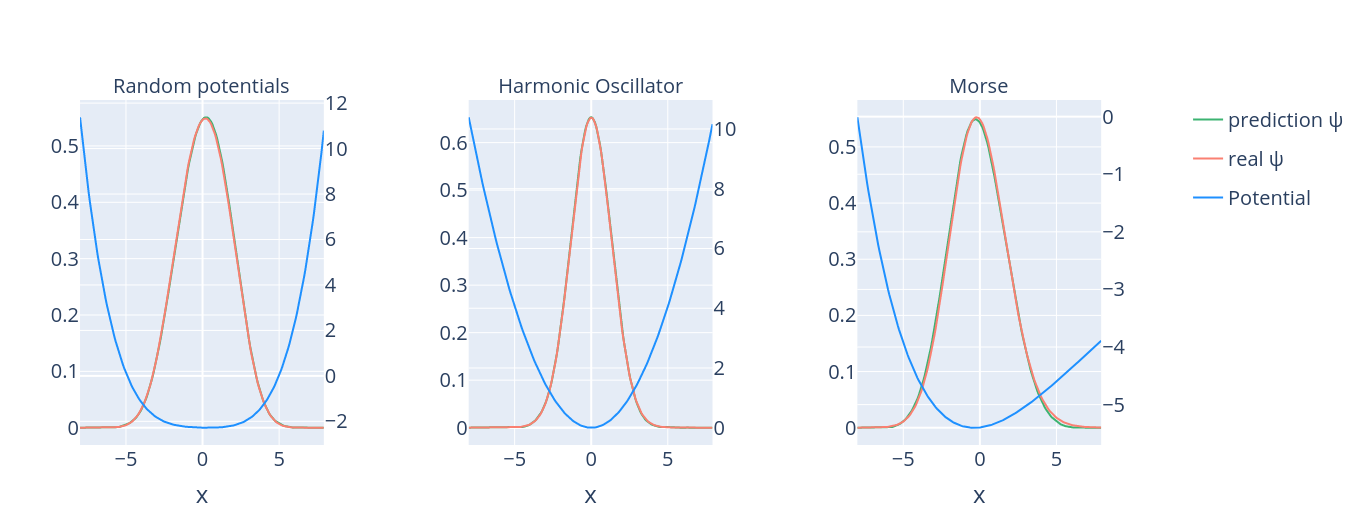}
 \caption{Example of the neural network prediction for three different potentials: random polynomial potential (left), 
 harmonic oscillator (middle), and Morse potential (right). 
 Each plot displays the potential (blue), the prediction of the fundamental eigenfunction (green) and the true eigenfunction (orange). }
 \label{fig:3}
\end{figure}

\begin{figure}
\centering
  \includegraphics[width=0.85\columnwidth]{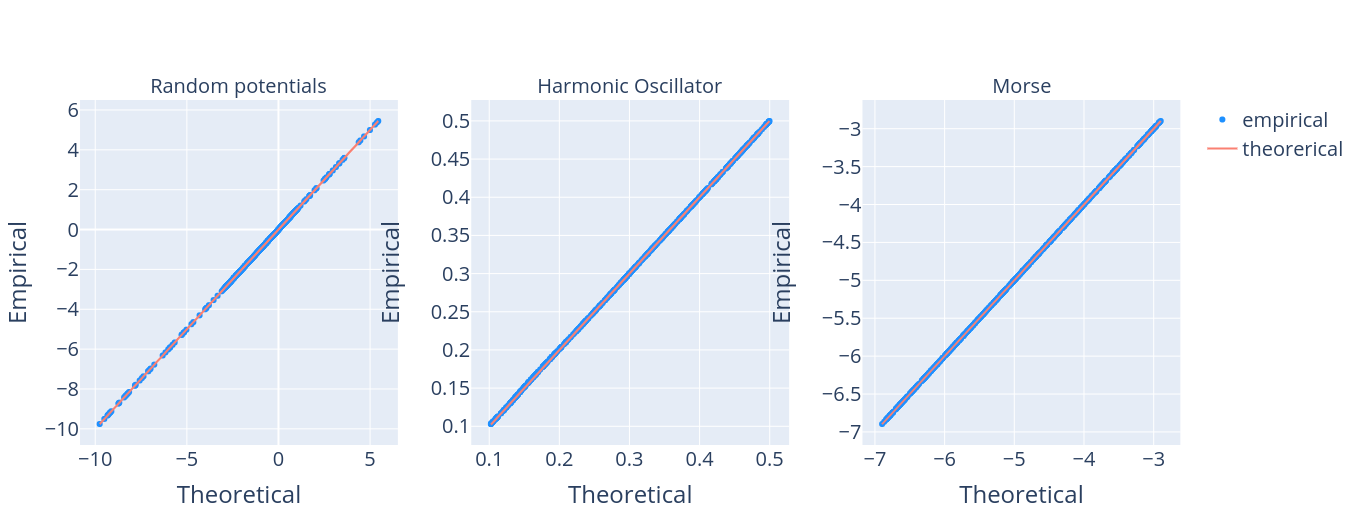}
 \caption{Predicted fundamental mean energies for three types of potentials: random polynomial potential (left), 
 harmonic oscillator (middle), and Morse potential (right). Some 500 samples are shown for each type of potential. }
 \label{fig:4}
\end{figure}

\begin{figure}
\centering
  \includegraphics[width=0.85\columnwidth]{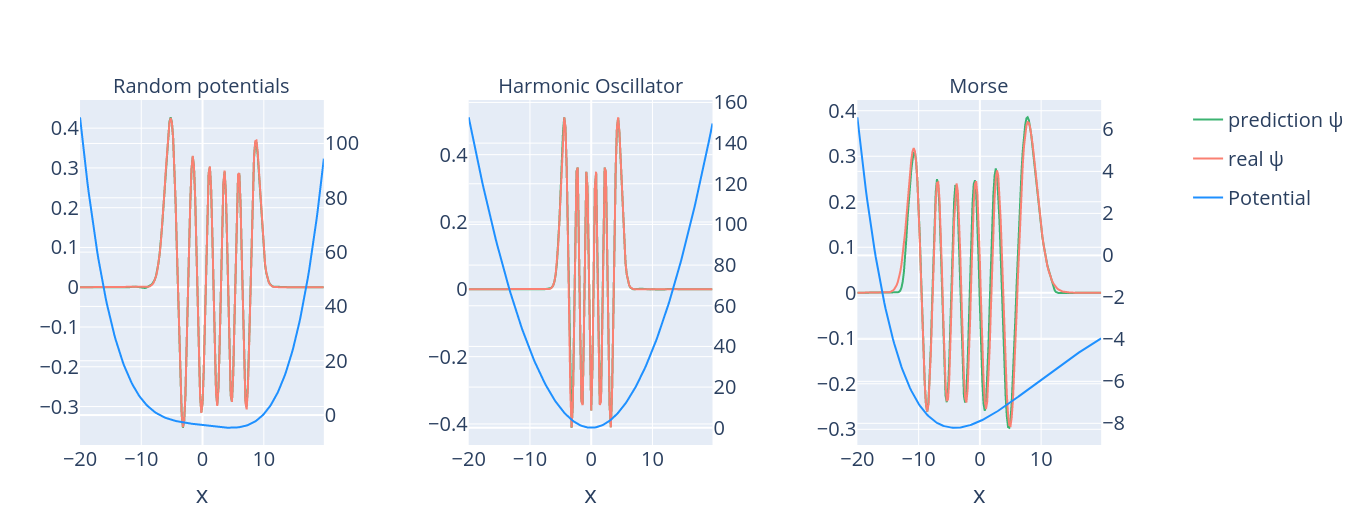}
 \caption{Example of the neural network prediction for three different potentials: random polynomial potenital (left), 
 harmonic oscillator (middle) and Morse potential (right). Each plot displays the potential (blue), the prediction of the 10-th 
 excited eigenfunction (green) and the true eigenfunction (orange). }
 \label{fig:5}
\end{figure}

\begin{figure}
\centering
  \includegraphics[width=0.85\columnwidth]{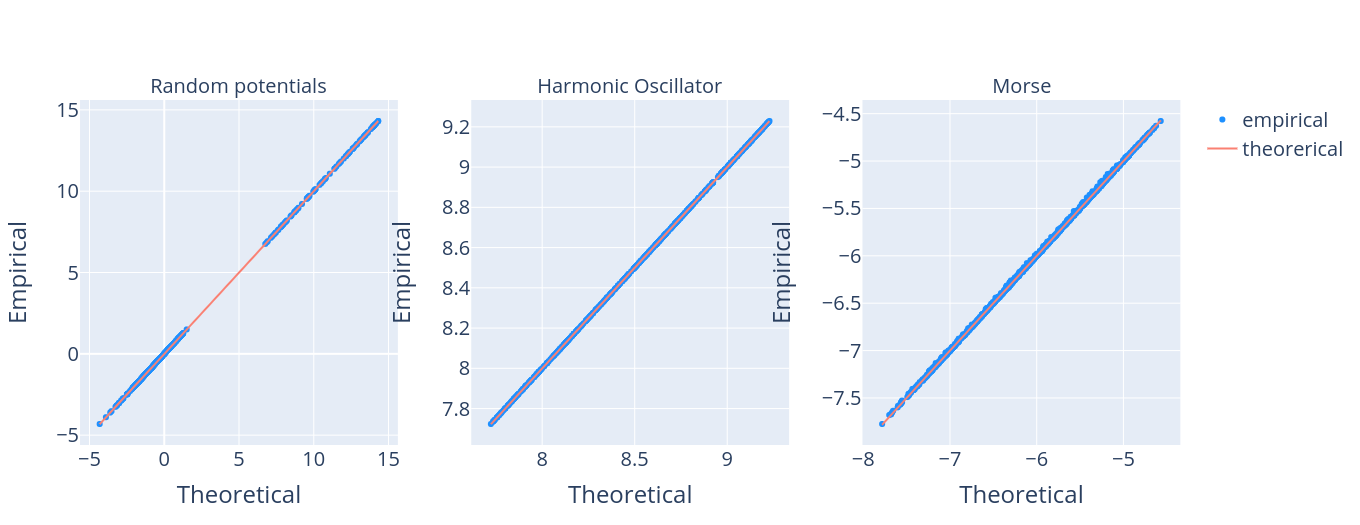}
 \caption{Predicted mean energies of the 10th excited eigen state for three types of potentials: random polynomial potenital (left), 
 harmonic oscillator (middle), and Morse potential (right). Some 500 samples are shown for each type of potential. }
 \label{fig:6}
\end{figure}

Two NNs were trained for this purpose, 
the first one to reproduce the fundamental eigenfunction, 
and the second to reproduce excited eigenfunctions, which in this case was chosen as that corresponding to $n=10$. 
Table~\ref{table:2} shows the mean square error (MSE) obtained for the predicted wavefunctions values and also for the mean energies 
of such vibrational states. 
These results are also shown graphically in Figs.~\ref{fig:3} and \ref{fig:4} for the fundamental state, 
and in Figs.~\ref{fig:5} and \ref{fig:6} for the excited state. 

In the case of the Morse Hamiltonian ground state, the MSE for both energy and wavefunction is similar to those 
obtained for the polynomial potentials. 
This fact means that the NN can effectively generalize to non-polynomial potentials when trained with polynomial potentials. 
The MSE of the harmonic oscillator potentials is also similar to the MSE of the random potentials, 
which is not an unexpected result since the harmonic oscillator is also a polynomial potential as well. 
Finally, it should be remarked that, since the MSE of the wave function is similar for all potentials, the NN is not producing much overfitting.

Regarding the excited eigenfunctions, the MSE for both the energy and wave function is small for the three types of potentials. 
However, we observe that the wave function prediction for the Morse potentials presents higher MSE than the MSE of the polynomial potentials. 
In particular, Fig.~\ref{fig:5} shows that the tails of the wave function are not correctly reproduced. 
This is a consequence of training the NN only with polynomial potentials, whose wave functions have a significantly different decay. 
However, we see that this error in the tails of the wave function does not affect much the value of the mean energy, 
since the MSE for the energy is similar to the MSE of the polynomial potentials. 
%
\begin{table}
\begin{tabular}{lll}
\hline
Potential Type       & MSE($\psi$)         & MSE(E)\\
\hline
Polynomial potentials & $6\cdot 10^{-6}$ & $6\cdot 10^{-8}$\\
Harmonic oscillator   & $3\cdot 10^{-5}$ & $3 \cdot 10^ {-6}$\\
Morse potential s     & $1\cdot 10^{-5}$ & $2 \cdot 10^{-6}$\\
\hline
\end{tabular}
\qquad
\begin{tabular}{lll}
\hline
Potential Type         & MSE($\psi$)        & MSE(E)\\
\hline
Polynomial potentials & $3\cdot 10^{-5}$ & $1 \cdot 10^{-6}$ \\
Harmonic oscillator   & $2\cdot 10^{-6}$ & $2 \cdot 10^{-7}$ \\
Morse potentials      & $6\cdot 10^{-3}$ & $9 \cdot 10^{-6}$\\
\hline
\end{tabular}
\caption{Mean square error (MSE) for the wavefunctions and energies for three types of potentials for the fundamental state (left) 
and the 10-th excited state (right).}
\label{table:2}
\end{table}
%
%
%
%
%

\begin{figure}[h!]
\centering
  \includegraphics[width=0.85\columnwidth]{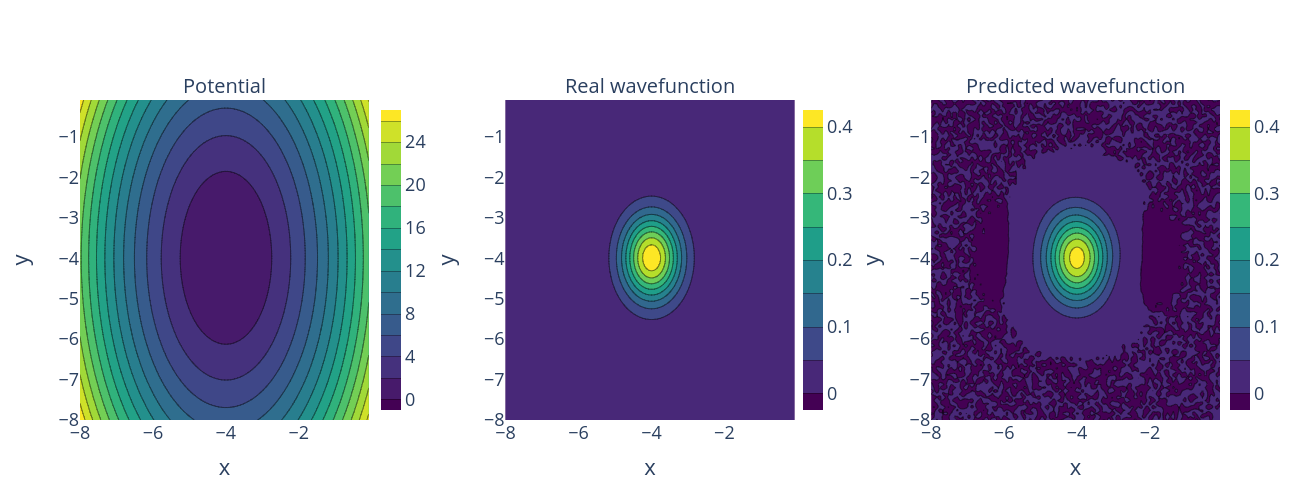}
 \caption{Example of potential (left), true eigen function (middle) and predicted eigen function (right) for a harmonic oscillator Hamiltonian. }
 \label{fig:7}
\end{figure}

\begin{figure}[h!]
\centering
  \includegraphics[width=0.85\columnwidth]{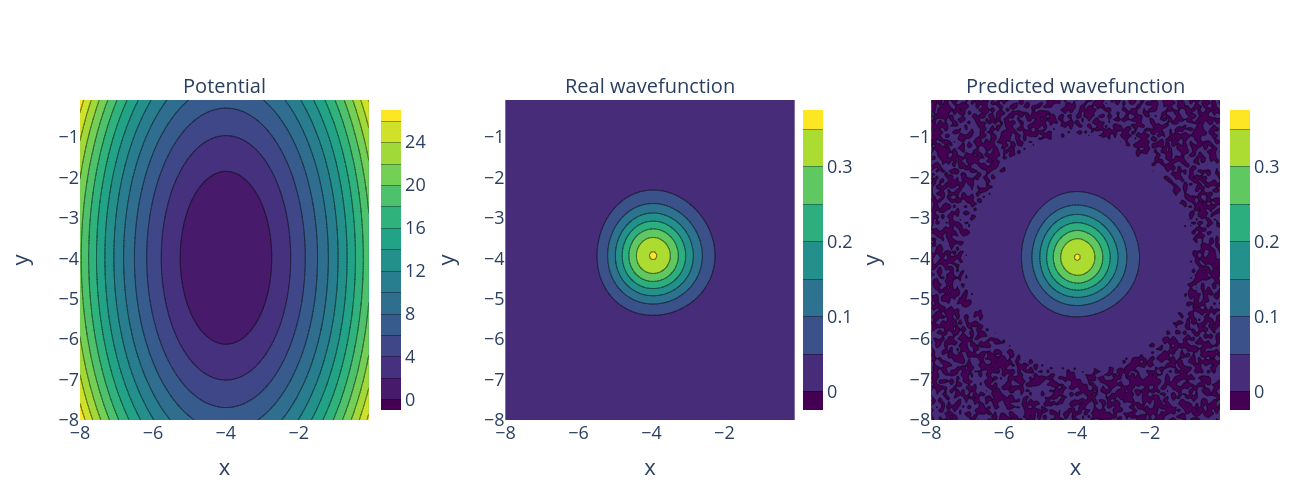}
 \caption{Example of potential (left), true eigenfunction (middle) and predicted eigenfunction (right) for a Morse potential. }
 \label{fig:8}
\end{figure}

\begin{figure}[h!]
\centering
  \includegraphics[width=0.85\columnwidth]{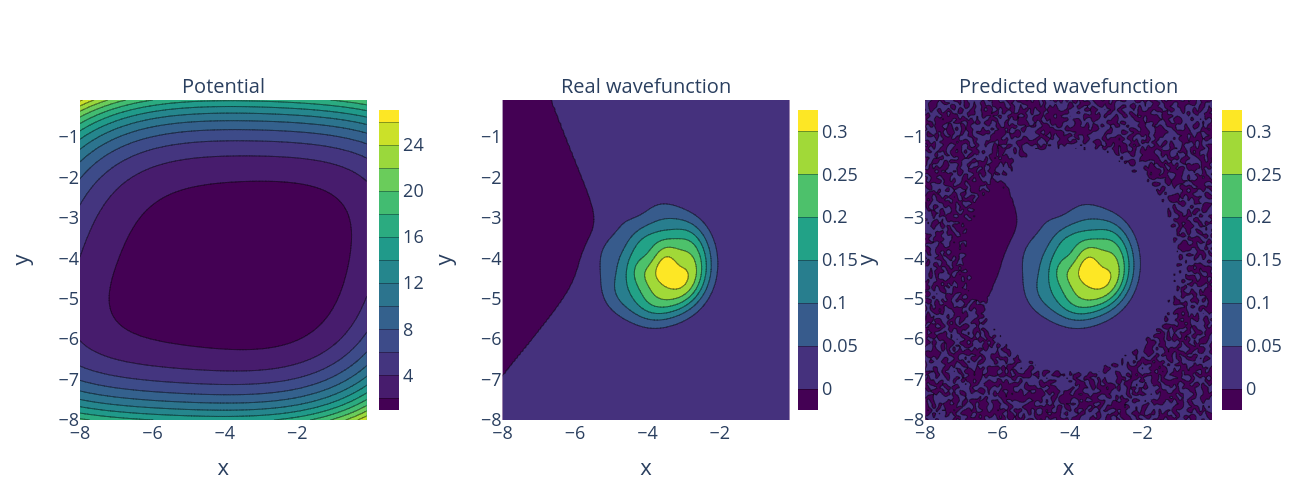}
 \caption{Example of potential (left), true eigenfunction (middle) and predicted eigenfunction (right) for a random polynomial potential. }
 \label{fig:9}
\end{figure}

\begin{figure}[h!]
\centering
  \includegraphics[width=0.85\columnwidth]{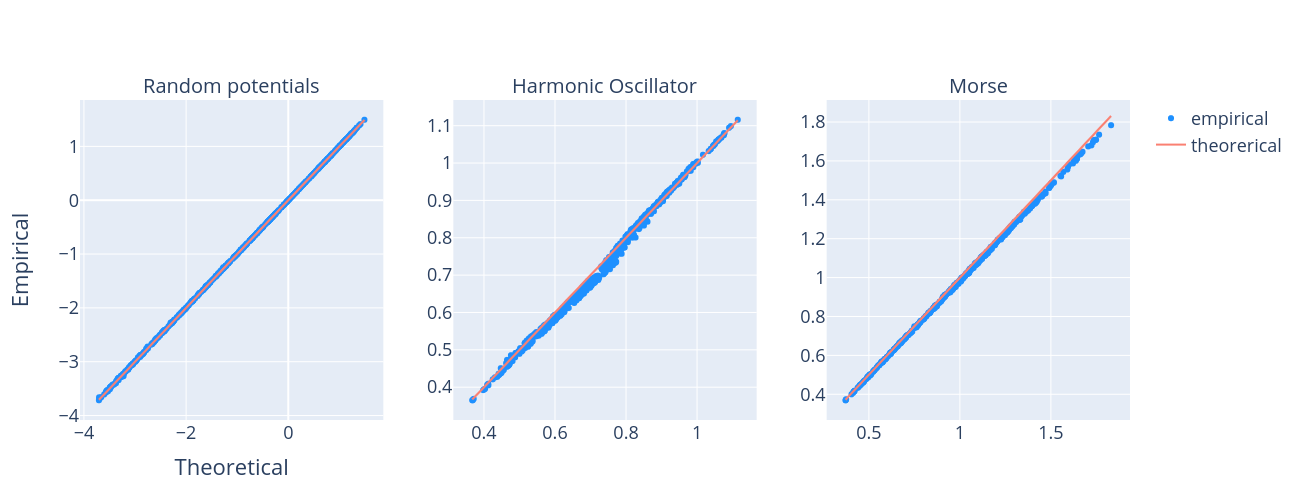}
 \caption{Predicted mean energies of the fundamental eigen state for three types of potentials: random polynomial potenital (left), 
 harmonic oscillator (middle) and Morse potential (right). Some 500 samples are shown for each type of potential. }
 \label{fig:10}
\end{figure}

A different NN was trained to reproduce the fundamental wave function of several 2D (harmonic, Morse and random) potentials. 
Again, the training data consists of random polynomial potentials, while the test data contains also decoupled Morse potentials, 
see Eq.~(\ref{eq:13}). 
The MSE results for the three cases are summarized in Table \ref{table:3}. 
Figures \ref{fig:7}, \ref{fig:8} and \ref{fig:9} show examples of the potential, the real eigenfunction and the predicted eigenfunction 
for the three types of potentials, and Fig.~\ref{fig:10} shows the corresponding the mean energy for the three types of potentials. 
We observe that the network is also able to reproduce the fundamental wave functions for 2D Morse potentials, 
as the MSE for both the energy and the wave functions is similar to the MSE of the random potentials. 
\begin{table}
\begin{tabular}{lll}
\hline
Potential Type         & MSE($\psi$)        & MSE(E)\\
\hline
Polynomial potentials & $2\cdot 10^{-7}$ & $5\cdot 10^{-5}$   \\
Harmonic oscillator   & $9\cdot 10^{-6}$ & $2 \cdot 10^ {-4}$ \\
Morse potentials     & $9\cdot 10^{-6}$ & $3 \cdot 10^{-4}$   \\
\hline
\end{tabular}
\caption{Mean square error for the eigen function and energies for three types of 2D potentials for the fundamental state.}
\label{table:3}
\end{table}
%
%
%
%

\subsection{Model II: Morse potentials}
  \label{sec:res-morse}

\begin{figure}[h!]
\centering
  \includegraphics[width=0.85\columnwidth]{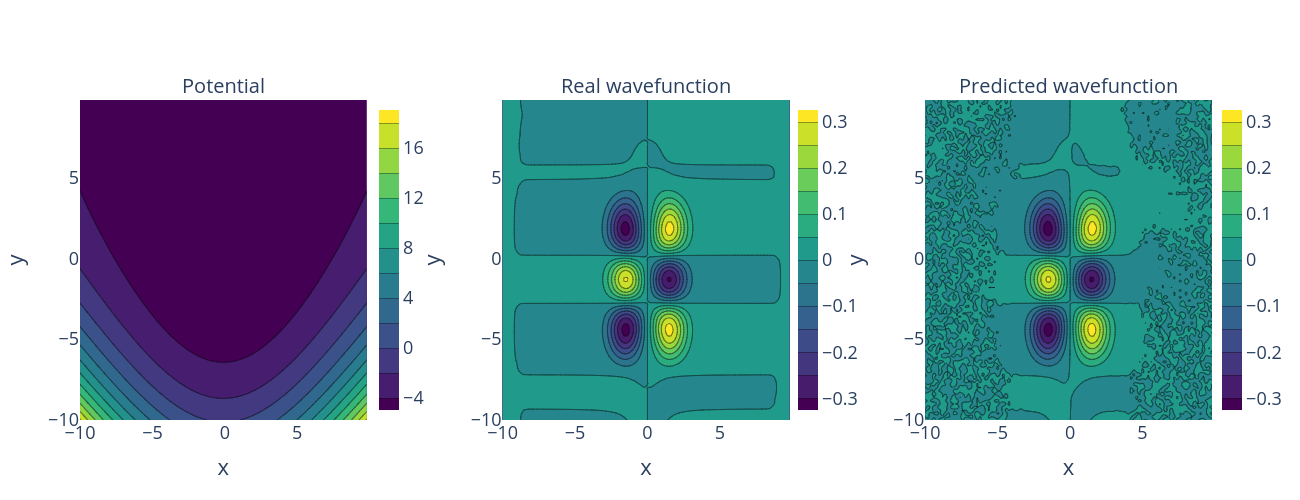}
 \caption{Example of potential (left), true eigen function (middle) and predicted eigen function (right) for the excited 
 eigenfunction with quantum numbers $n_x =1$, $n_y=2$. The Morse parameters are $D_e = 2.5$, $a=0.095$. }
 \label{fig:11}
\end{figure}

\begin{figure}[h!]
\centering
  \includegraphics[width=0.85\columnwidth]{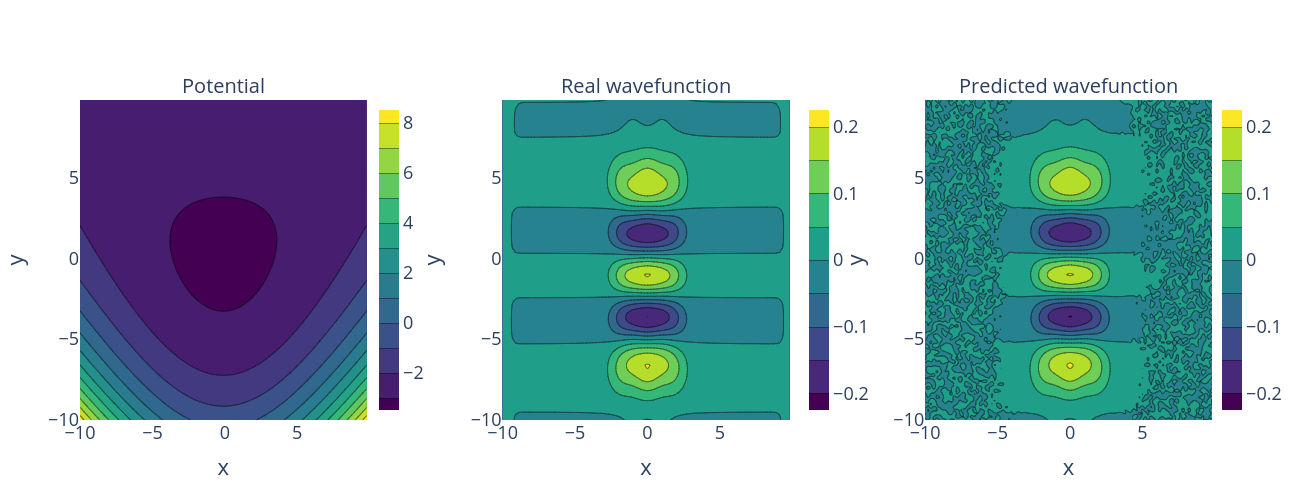}
 \caption{Example of potential (left), true eigen function (middle) and predicted eigen function (right) for the excited 
 eigenfunction with quantum numbers $n_x =0$, $n_y=5$. The Morse parameters are $D_e = 2.1$, $a=0.095$. }
 \label{fig:12}
\end{figure}

\begin{figure}[h!]
\centering
  \includegraphics[width=0.85\columnwidth]{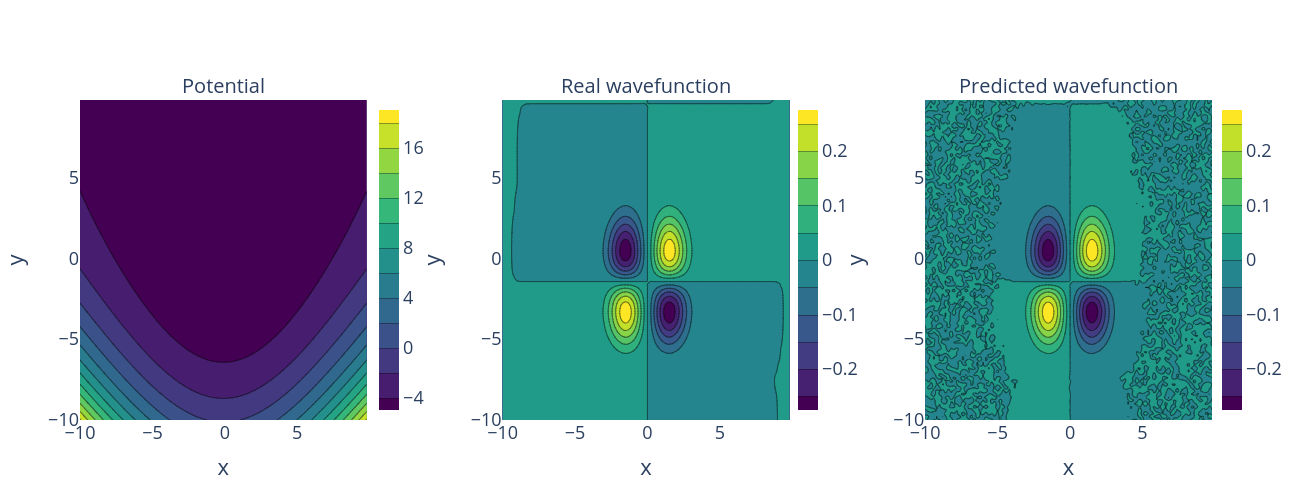}
 \caption{Example of potential (left), true eigen function (middle) and predicted eigen function (right) for the excited 
 eigenfunction with quantum numbers $n_x =1$, $n_y=1$. The Morse parameters are $D_e = 2.4$, $a=0.097$. }
 \label{fig:13}
\end{figure}

\begin{figure}[h!]
\centering
  \includegraphics[width=0.85\columnwidth]{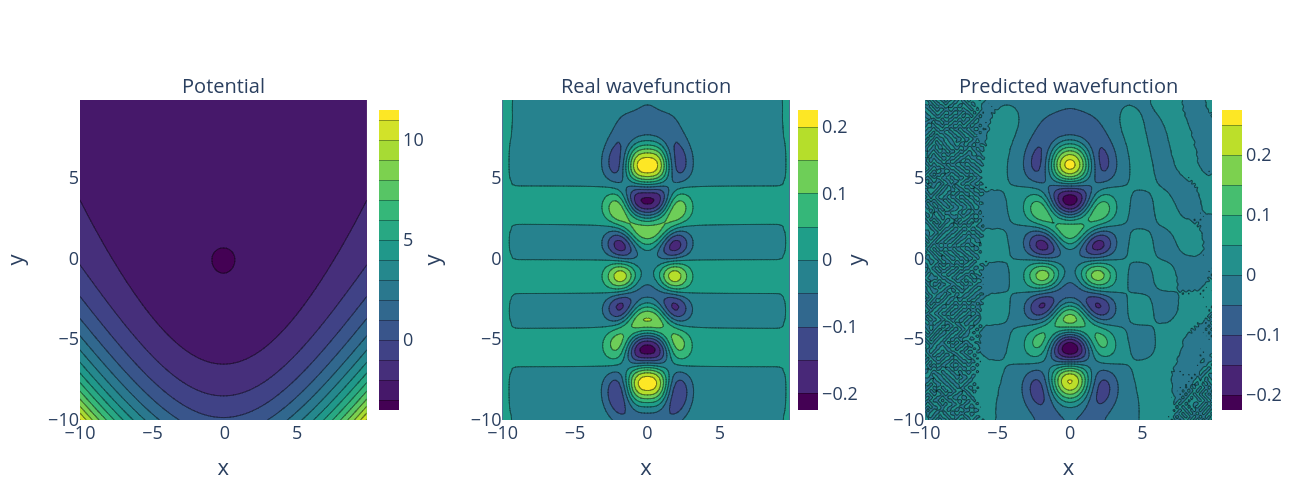}
 \caption{Example of potential (left), true eigen function (middle) and predicted eigen function (right) for the excited 
 eigenfunction with quantum numbers $n_x =1$, $n_y=2$. The Morse parameters are $D_e = 1.6$, $a=0.098$. }
 \label{fig:14}
\end{figure}

\begin{figure}[h!]
\centering
  \includegraphics[width=0.85\columnwidth]{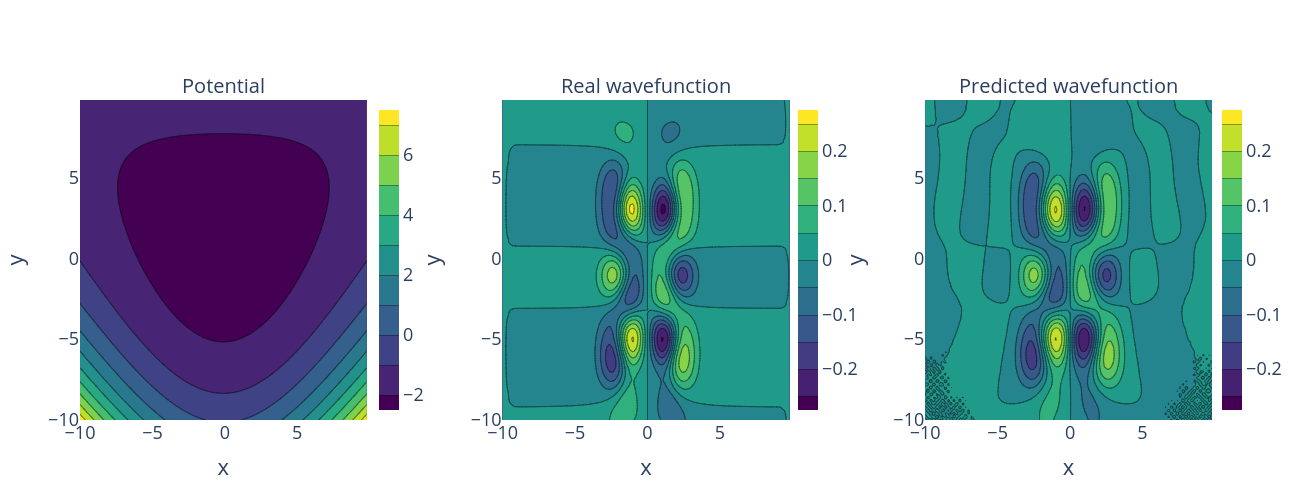}
 \caption{Example of potential (left), true eigen function (middle) and predicted eigen function (right) for the excited 
 eigenfunction with quantum numbers $n_x =2$, $n_y=3$. The Morse parameters are $D_e = 1.1$, $a=0.093$. }
 \label{fig:15}
\end{figure}

\begin{figure}[h!]
\centering
  \includegraphics[width=0.85\columnwidth]{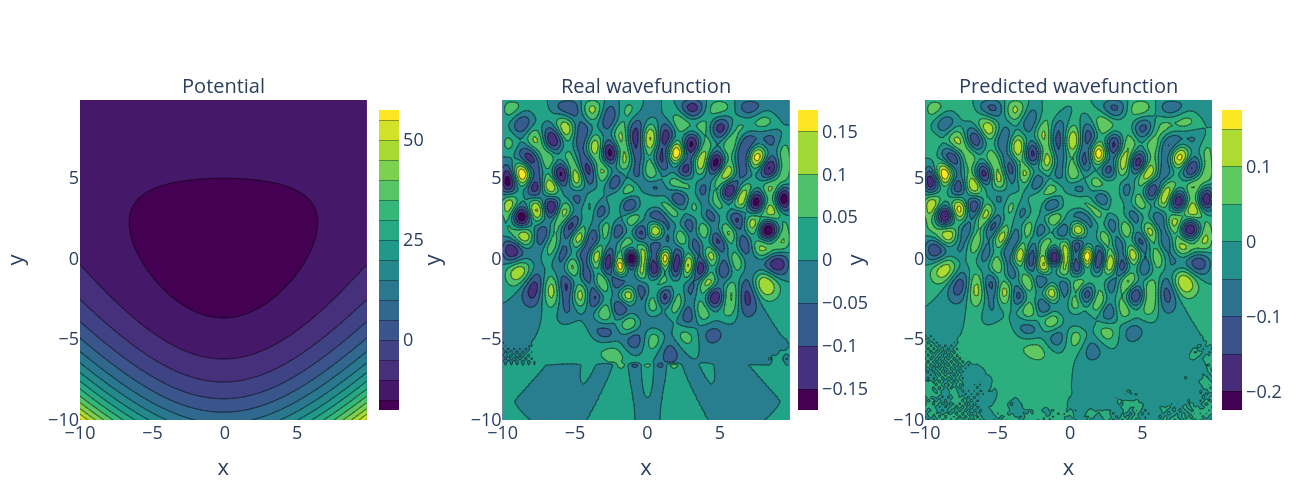}
 \caption{Example of potential (left), true eigen function (middle) and predicted eigen function (right) for a high-energy 
 eigenfunction. The Morse parameters are $D_e = 7.8$, $a=0.098$, $G_{12}=0.35$. }
 \label{fig:16}
\end{figure}

\begin{figure}[h!]
\centering
  \includegraphics[width=0.85\columnwidth]{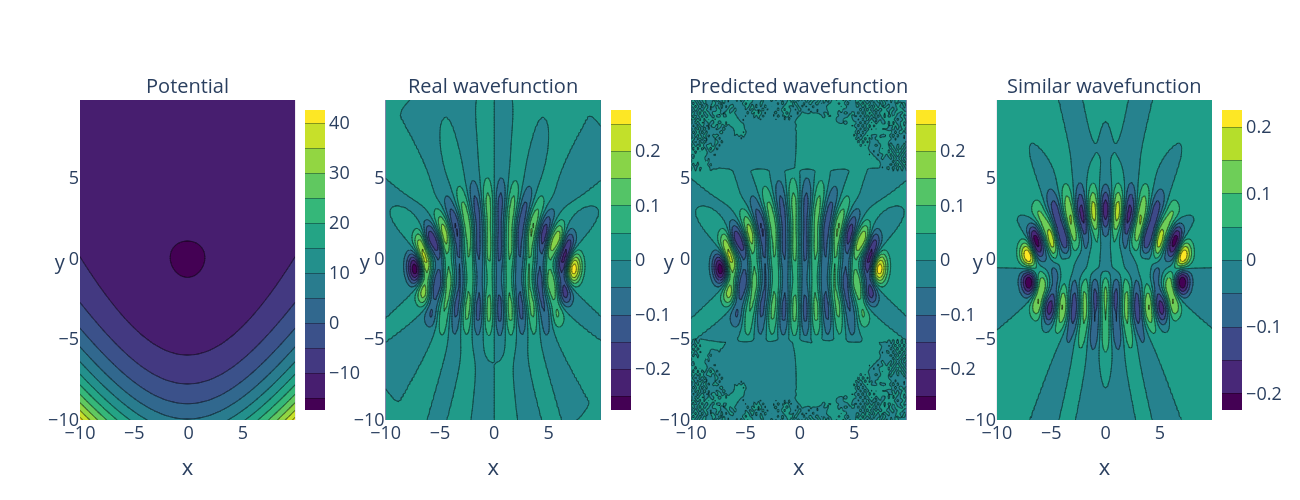}
 \caption{Example of potential (first), true eigen function (second) and predicted eigen function (third) for the excited 
 eigenfunction with quantum numbers $n_x =17$, $n_y=1$. The fourth plot displays a similiar eigenfunction to the predicted one. 
 The Morse parameters are $D_e = 7.1$, $a=0.096$, $G_{12}=0.35$. }
 \label{fig:17}
\end{figure}

In the second part of this work our goal is to design a NN model which is able to obtain the excited states of more
realistic, yet complex, molecular potentials. 
In particular, we train a NN using a separable Hamiltonian containing Morse potentials in both spatial directions. 
These Hamiltonians have analytical solutions for the Schr\"{o}dinger equation, and thus no numerical solver is needed. 
Afterwards, the NN is asked to generate some excited eigenfunctions for a coupled version of the previous Morse potentials. 
The results are shown in Figs.~\ref{fig:11}-\ref{fig:17}. 
%
%
%
%
%
%
%

As can be seen, for low energies, the eigenfunctions of the coupled Morse potential exhibit a well-defined nodal pattern which leads
to an unambiguous quantum numbers assignment; see, for example, Figs.~\ref{fig:11}, \ref{fig:12}, \ref{fig:13}, 
which correspond to states $(n_x,n_y)=(1,2), (0,4)$, and $(1,1)$, respectively. 
These wave functions are very similar in topology to the eigenfunctions of the decoupled Morse potential, 
which is used to train the NN. 
Therefore, the NN can easily generate the eigenfunctions for the coupled Morse potential. 
In this case, the strategies used to generate the training data are enough to make the NN able to generate the 
eigenfunctions for coupled Morse potentials. 
However, for higher energy levels; see Figs.~\ref{fig:14} and \ref{fig:15}, the eigenfunctions present sizeable distortions with
respect to the uncoupled ones, and the quantum numbers can not be defined so easily. 
In these cases, it is necessary to add the loss function described in section \ref{sec:NN_coupled} 
to make the NN able of predicting this kind of wavefunctions. 
In order to increase the distortion of the eigenfunctions for the coupled Morse potential and make them more challenging for being 
found by our NN, we artificially increase the coupling factor $G_{12}$ to $G_{12}=0.35$. 
With this new value of the coupling parameter, high energy states with a high distortion were also reproduced; 
two representatives examples are shown in Figs.~\ref{fig:16} and \ref{fig:17}. 
Figure~\ref{fig:17} contains an additional plot, showing a wave function with similar energy and similar quantum numbers. 
When the energy levels increase, there may be more than one wavefunction with a very close energy and similar quantum numbers. 
In these cases, using the eigenfunctions of decoupled Morse potentials to train the network only allows to reproduce one of the wave functions, 
i.e.,~that which is more similar to the decoupled wave function. 
If we wanted to distinguish between these eigenstates, we would have to provide a finer approximation to the target wave function. 
One option is to gradually increase the coupling factor $G_{12}$ to approach the final desired Hamiltonian, using at each
step in the approximation the coupled Morse potential with for the previous (smaller) coupling factor 
to approximate the wavefunctions corresponding to higher-coupled potentials. 
This technique will be explored in a future work.   

All in all, our results demonstrate that the method that we propose allows to reproduce a great variety of wave functions,
at least for systems similar to that of the coupled Morse potentials, using only the wavefunctions of the decoupled,
yet separable, models, which in the case of the Morse potential have an analytical solution.

\section{Summary and conclusions}
  \label{sec:summary}

DNNs have proven to be very useful for multiple technological applications of everyday life
\citep{Application1,Application2, Application3, Application4}. 
In particular, CNNs are capable of extracting features from spatial data that are useful 
for the learning task \cite{CNN}. 
In this project, we use the paradigm of DNNs to produce the fundamental and excited eigenfunctions 
of molecular Hamiltonians.

In the first part of the paper, we train a FCNN (1D) and a CNN (2D) using randomly-generated polynomial potentials 
and their associated eigenfunctions. 
Then, the network is asked to generalized to unseen, non-polynomial potentials, in particular, 
we used the well known Morse potentials for the test set. 
The obtained results show that even though the network was only trained to generate polynomial potentials, 
it was able to correctly reproduce the fundamental and excited eigenfunctions for different Morse potentials. 
However, the tails of the wave function were not always reproduced totally correct.  
This is a consequence of training the network only with polynomial potentials, which have a significant different 
asymptotic decay.
In any case, this is not a terrible result, since the importance of these tails in the computation of
molecular properties is usually very limited. 
For example, these errors do not affect much the value of the mean energy of the system. 
Moreover, our results show the advantages of using ML instead of a model-based approach.
Even though the data generation process was very different for the polynomial (training) and Morse potentials
(prediction), the model was able to reproduce the Morse potential wave functions. 
 
The second goal of the paper is more challenging. 
Given a zeroth-order Hamiltonian $H_0$ with known eigenfunctions, 
we aim to obtain the eigenfunctions of $H$, assuming that $H_0$ is an approximation of $H$. 
We applied this framework, which resembles that of perturbation theory in the Kolmogorov-Arnold-Moser scenario,
to find the eigenfunctions of a well known 2D coupled Morse model, which represents well the stretching 
dynamics of the H$_2$O molecule. 
A CNN was trained using the potential and eigenfunction of multiple decoupled Morse Hamiltonians. 
Such Hamiltonians are separable and their eigenfunctions are well-known analytically. 
Therefore, they are very suitable to use as a training data set, since no numerical integration is needed to generate the data. 
The data generation process and learning algorithm were crucial for guaranteeing the good performance of the algorithm. 
The former consisted in choosing the appropriate Morse parameters and the $(x,y)$ domain so that 
the training data was as similar as possible to the test data. 
Regarding the learning algorithm, we introduced an additional loss function which ensured that the predicted 
wavefunction was an eigenstate of the coupled Morse Hamiltonian, and thus a solution of our problem. 
By using all these techniques we have been able to reproduce high-energy states of coupled Morse potentials. 
Obtaining these wavefunctions was challenging since they presented high distortions with respect to the
training (zeroth order) states, and the quantum numbers were not easily defined. 
Therefore, our results prove that DNNs trained with the appropriate learning algorithm can 
reproduce high-energy eigenstates of complex Hamiltonians. 
This can be considered as a good first step towards the computation of vibrational wavefunctions in high
dimensional systems, where the ML methods bear a clear advantage over the standard computational
chemistry approaches. 

\section*{Declaration of Competing Interest}
The authors declare no competing interest. 
\section*{Code availability}
All codes and data used for the development of the project associated with the current submission are
available at \href{https://github.com/laiadc/DL-schrodinger}{https://github.com/laiadc/DL-schrodinger}. 
Any updates will also be published on GitHub.

%
\section*{Acknowledgments}
The project that gave rise to these results received the support of a fellowship from "la Caixa" Foundation (ID 100010434). 
The fellowship code is LCF/BQ/DR20/11790028.
This work has also been partially supported by the Spanish Ministry of Science, Innovation and Universities, 
Gobierno de Espa\~na, under Contracts No.\ PGC2018-093854-BI00, ICMAT Severo Ochoa CEX2019-000904-S;
and by the People Programme (Marie Curie Actions) of the European Union's Horizon 
2020 Research and Innovation Program under Grant No.\ 734557.
%
\newpage
\appendix
\section{The Variational method for a harmonic oscillator basis set}
  \label{sec:appendixA}
Given the 1D Hamiltonian
\begin{equation}
\hat{H} = \frac{\hat{p}^2}{2m} + V(\hat{x}),
\end{equation}
the mean energy corresponding to the normalized wavefunction $\psi(x)$ is given by
\begin{equation}
\expval{H} = \expval{H}{\psi} = \int_{-\infty}^\infty \psi^*(x) H(x) \psi(x)dx.
\end{equation}
For simplicity, we will use from now on $m=1$ and $\hbar=1$ 
(which are the values that have been used throughout this work) 
and thus, these parameters will be omitted in all expressions and calculations.

\subsection*{Variational principle}
The variational principle states that the mean energy under a Hamiltonian $H$ for a wave function is always 
greater or equal to the exact ground state energy of such Hamiltonian. That is
\begin{equation}
E_0 \leq \expval{H} = \expval{H}{\psi} \quad \forall \ket{\psi} \in \mathcal{H}.
\end{equation}
This principle can be extended to higher eigenenergies by imposing that the state $\psi$ is orthogonal to the 
previous eigenstates.

\subsection*{The harmonic oscillator basis set}

We choose as a basis set for $\mathcal{H}$ consisting of the eigenfunctions of the harmonic oscillator (HO) 
with $\omega=1$, i.e., 
\begin{equation}
\phi_n(x) = \frac{1}{\sqrt{2^n n! \sqrt{\pi}}} e^{-x^2/2} H_n(x)\, ,
\end{equation}
where $H_n(x)$ is the $n$-th Hermite polynomial, defined by the recurrence
\begin{equation}
H_n(x) = 2xH_{n-1}(x) - 2n H_{n-2}(x),\qquad \mbox{and} \quad H_0(x) = 1, \ H_1(x) = x^2.
\end{equation}

Since $\{\phi_n(x)\}_n$ form a complete basis for $\mathcal{H}$, we can write any wavefunction $\psi(x)$ as a 
linear combination of the eigenfunctions or basis set elements$\{\phi_n\}$.
\begin{equation}
\psi(x) = \sum_{n=0}^\infty a_n \phi_n(x),
\end{equation}
and the associated mean energy of $\psi$ is
\begin{multline}
\expval{H} = \expval{H}{\psi} = \int_{-\infty}^\infty \Big(\sum_{n=0}^\infty a_n \phi_n(x)\Big) \hat{H} 
  \Big(\sum_{m=0}^\infty a_m \phi_m(x)\Big) dx =\\
  \sum_{n=0}^\infty \sum_{m=0}^\infty a_n a_m \int_{-\infty}^\infty \phi_n(x) H(x) \phi_m(x)dx =
  \sum_{n=0}^\infty \sum_{m=0}^\infty a_n a_m C_{nm},
\end{multline}
where the coefficients $C_{nm}$ are given by
\begin{equation}
C_{nm} = \int_{-\infty}^\infty A_n e^{-x^2/2} H_n(x) \Big(-\frac{1}{2} \frac{\partial^2}{\partial x^2} 
   + V(x) \Big) A_m e^{-x^2/2} H_m(x) dx, \qquad \mbox{being} \quad A_n = \frac{1}{\sqrt{n!2^n \sqrt{\pi}}}.
\end{equation}
Finally, in order to find a good estimation of the ground state, one uses a finite basis of HO eigenstates 
$\{\phi_n\}_{n=0}^N$ and then find the coefficients $\{a_n\}$ which minimize the mean energy $\expval{H}$. 

\subsection*{Finding the coefficients $\{a_n\}$}

To find the coefficients $\{a_n\}$ which minimize the energy $\expval{H}$ we use the Lagrange Multipliers theorem, 
which state that the local minimum of a function $F$, under a constraint $G$ is the solution of
\begin{equation}
\nabla F = \lambda \nabla G, \quad \lambda \in \mathbb{R}.
\end{equation}
In this case $F$ is the mean energy $F(\{a_n\}) = \expval{H}$, and $G$ is the normalization constraint. 
Since $\{\phi_n\}$ is a basis set of of $\mathcal{H}$, then $G(\{a_n\}) = \sum_{n=0}^N a_n^2 = 1$.

Next, we calculate the partial derivative of $F$
\begin{multline}
\frac{\partial F}{\partial a_i} = \frac{\partial}{\partial a_i} \Big( \sum_{n=0}^N \sum_{m=0}^N a_n a_m C_{nm} \Big)
=  \frac{\partial}{\partial a_i} \Big( \sum_{n=0}^N a_n \Big) \Big(\sum_{m=0}^N a_m C_{nm}\Big) 
+ \frac{\partial}{\partial a_i} \Big( \sum_{m=0}^N a_m \Big) \Big(\sum_{n=0}^N a_n C_{nm}\Big) =\\
\sum_{m=0}^N a_m C_{im} + \sum_{n=0}^N a_n C_{ni} = \sum_{n=0}^N a_n (C_{in} + C_{ni}),
\end{multline}
so that the gradient of $F$ is linear equations system
\begin{equation}
\nabla F(\vec{a}) = D \vec{a}, \quad \vec{a} = 
\begin{pmatrix} 
a_1\\
\vdots\\
a_N
\end{pmatrix}, \quad D \in \mathcal{M}_{N}(\mathbb{R}), \ [D]_{ij} = C_{ij} + C_{ji}.
\end{equation}
Moreover, the partial derivative of $G$ is
\begin{equation}
\frac{\partial G}{\partial a_i} = \frac{\partial }{\partial a_i} \Big(\sum_{n=0}^N a_n^2\Big) = 2a_i ,
\end{equation}
and the Lagrange Multiplier equation becomes
\begin{equation}
\nabla F(\vec{a}) = \lambda \nabla G(\vec{a}) \Longleftrightarrow D \vec{a} = 2\lambda \vec{a},
 \label{eq:41}
\end{equation}
which is an eigenvalue problem. 

The solution will then be found by solving the eigenvalue problem (\ref{eq:41}), and then selecting the vector $\vec{a_0}$ 
which minimizes $\expval{H}$. 
Since the basis $\phi_n(x)$ is finite (we take up to $N$ functions), the solution will be an approximation of the true eigenvector. 
When $N \rightarrow \infty$ the solution $\psi(x)$ will converge to the ground state of $H$. 
Moreover, since the eigenvectors of $D$ are orthonormal, the vector with the $n$th lowest energy will be an 
approximation to the $n$-th excited state of the Hamiltonian.  

\subsection*{Integrals involving Hermite polynomials}

In order to generate a basis of the Hilbert space that we will use to approximate the gound state wavefunctions,
we need to perform some integrals involving Hermite polynomials, i.e.,
\begin{equation}
I(n,m,r) = \int_{-\infty}^\infty x^r e^{-x^2} H_n(x) H_m(x) dx
\end{equation}
%
%
which can be obtained by the recurrence relation
\begin{multline}
I(n,m,r) = \int_{-\infty}^\infty x^r e^{-x^2} H_n(x) H_m(x) dx = 
  \int_{-\infty}^\infty x^r e^{-x^2}\frac{1}{2x}\Big(H_{n+1}(x) + 2nH_{n-1}(x)\Big)H_m(x)dx = \\
  \frac{1}{2}I(n+1,m,r-1) + nI(n-1,m,r-1), 
\end{multline}
and taking into account that
\begin{equation}
I(n,m,0) = \sqrt{\pi} 2^n n! \delta_{n,m} 
\end{equation}

\subsection*{Calculating $C_{nm}$}

In order to compute matrix $D$ in the eigenproblem expression (\ref{eq:41}), we need to calculate the coefficients $C_{nm}$
\begin{equation}
C_{nm} = A_nA_m \int_{-\infty}^\infty e^{-x^2/2} H_n(x) (-\frac{1}{2} \frac{\partial^2}{\partial x^2} 
+ V(x)) H_m(x) e^{-x^2/2} dx,
 \qquad \mbox{with} \quad A_n = \frac{1}{\sqrt{n! 2^n \sqrt{\pi}}}.
\end{equation}
In order to do so we need to calculate
\begin{equation}
\frac{\partial^2}{\partial x^2}(H_m(x) e^{-x^2/2} ) = e^{-x^2/2}\Big((x^2-1) H_m(x) - 4mx H_{m-1}(x) 
+ 4m(m-1)H_{m-2}(x)\Big) := e^{-x^2/2} P(x)
\end{equation}
\begin{multline}
C_{nm} = A_n A_m \Big( - \frac{1}{2} \int_{-\infty}^\infty  H_n(x) P(x) e^{-x^2} dx 
+ \int_{-\infty}^\infty e^{-x^2} H_n(x)H_m(x)V(x) dx\Big) = \\
A_nA_m\Big(- \frac{1}{2} I(n,m,2) + 1/2 I(n,m,0) + 2mI(n,m-1,1) - 2m(m-1)I(n, m-2, 0) + I_V\Big),
\end{multline}
where $I_V$ is the integral corresponding to the potential $V(x)$. If this potential is a polynomial
\begin{equation}
V(x) = \sum_{i=1}^N \alpha_i x^i ,
\end{equation}
then
\begin{equation}
I_V = \sum_{i=1}^N \alpha_i I(n,m,i) 
\end{equation}

\section{Variational method in 2D}
\label{sec:appendixB}
The previous variational method can be extended to 2D as described below (we will only focus on the differences between 
the 1D and 2D problems). 

\subsection*{The harmonic oscillator basis set for 2D}

We choose as a basis of $\mathcal{H}$ the eigenfunctions of the harmonic oscillator for 
$\omega_x=1$, $\omega_y=\omega \in \mathbb{R}-\mathbb{Q}$ in 2D, so that
\begin{equation}
\hat{H} = \frac{\hat{p}^2}{2m} + \frac{1}{2}m(x^2 + \omega^2y^2).
\end{equation}
Notice that the frequency is different for the two dimensions. 
Since this Schr\"{o}dinger equation is separable, the eigenfunctions come as the product of 1D eigenfunctions in both coordinates
\begin{equation}
\phi_{n_x, n_y}(x,y) = \frac{1}{\sqrt{n_x!2^{n_x} \sqrt{\pi}}}\frac{1}{\sqrt{n_y!2^{n_y} \sqrt{\pi/\omega}}} 
  e^{-x^2/2}e^{-y^2\omega/2} H_{n_x}(x)H_{n_y}(\sqrt{\omega}y) = \phi_{n_x}(x)\phi_{n_y,\omega}(y).
\end{equation}
%
%
and the corresponding eigenenergies come in terms of $n_x$ and $n_y$ as
\begin{equation}
E_{n_x, n_y} = \hbar (n_x + \omega n_y +1) .
\end{equation}
Since $\omega \in \mathbb{R}-\mathbb{Q}$ there is no degeneracy in the energy levels, 
which can then be ordered in an ascending mode. 
Therefore, there exists a unique bijective order in $(n_x,n_y)$ which sorts the energy levels. 
Hence we can write $\phi_{n_x,n_y}(x,y) = \phi_n (x,y)$ where $n = n (n_x,n_y)$. 
Since $\{\phi_n(x,y)\}_n$ form a complete basis set for $\mathcal{H}$, 
we can write any wavefunction $\psi(x,y)$ as the following linear combination
\begin{equation}
\psi(x,y) = \sum_{n=0}^\infty a_n \phi_n(x,y),
\end{equation}
and the associated mean energy is
\begin{multline}
\expval{H}= \expval{H}{\psi} = \int_{-\infty}^\infty\int_{-\infty}^\infty \Big(\sum_{n=0}^\infty a_n \phi_n(x,y)\Big) 
\hat{H} \Big(\sum_{m=0}^\infty a_m \phi_{m,\omega}(x,y)\Big) dx dy=\\
\sum_{n=0}^\infty \sum_{m=0}^\infty a_n a_m\int_{-\infty}^\infty \int_{-\infty}^\infty \phi_n(x,y) H(x,y) \phi_{m,\omega}(x,y)dx 
= \sum_{n=0}^\infty \sum_{m=0}^\infty a_n a_m C_{nm},
\end{multline}
where
\begin{multline}
C_{nm} = \int_{-\infty}^\infty \int_{-\infty}^\infty A_n e^{-x^2/2} H_{n_x}(x)e^{-y^2\omega/2} H_{n_y}(\sqrt{\omega}y) 
\Big(-\frac{1}{2} \frac{\partial^2}{\partial x^2} -\frac{1}{2} \frac{\partial^2}{\partial y^2} 
+ V(x,y) \Big)\cdot \\ A_m e^{-x^2/2}e^{-y^2\omega/2} H_{m_x}(x) H_{m_y}(\sqrt{\omega}y) dx dy, 
\qquad \mbox{with} \quad A_n = \frac{1}{\sqrt{n_x!2^{n_x} \sqrt{\pi}}}, \ A_m = \frac{1}{\sqrt{n_y!2^{n_y} \sqrt{\pi/\omega}}}
\end{multline}

\subsection*{Calculating $C_{nm}$}

Now, for any given potential
\begin{equation}
V(x,y) = \sum_{j+j\leq k} \alpha_{ij} x^i y^j ,
\end{equation}
and taking into account
\begin{equation}
\frac{\partial^2}{\partial x^2}(H_m(x) e^{-x^2/2} ) = e^{-x^2/2}\Big((x^2-1) H_m(x) - 4mx H_{m-1}(x) 
+ 4m(m-1)H_{m-2}(x)\Big) := e^{-x^2/2} P_m(x),
\end{equation}
and defining the new variable $\tilde{y} = \sqrt{\omega}y$, the expression for $C_{nm}$ becomes for 2D
\begin{multline}
C_{nm} = \int_{-\infty}^\infty \int_{-\infty}^\infty A_n e^{-x^2/2} H_{n_x}(x)e^{-\tilde{y}^2/2} H_{n_y}(\tilde{y}) 
\Big(-\frac{1}{2} \frac{\partial^2}{\partial x^2} -\frac{\omega}{2} \frac{\partial^2}{\partial \tilde{y}^2} 
+ V(x,\tilde{y}/\sqrt{\omega}) \Big)\cdot \\ A_m e^{-x^2/2}e^{-\tilde{y}^2/2} H_{m_x}(x) H_{m_y}(\tilde{y}) dx d\tilde{y}.
\end{multline}
Or alternatively
\begin{multline}
C_{nm} = A_n A_m \Big( - \frac{1}{2} \int_{-\infty}^\infty e^{-\tilde{y}^2} H_{n_y}(\tilde{y}) H_{m_y}(\tilde{y}) d\tilde{y}
\int_{-\infty}^\infty  H_{n_x}(x) P_{m_x}(x) e^{-x^2} dx - \\
\frac{\omega}{2} \int_{-\infty}^\infty e^{-x^2} H_{n_x}(x) H_{m_x}(x) dx\int_{-\infty}^\infty  
H_{n_y}(\tilde{y}) P_{m_y}(\tilde{y}) e^{-\tilde{y}^2} d\tilde{y} \ + \\
 \int_{-\infty}^\infty \int_{-\infty}^\infty e^{-x^2-\tilde{y}^2} H_{n_x}(x)H_{n_y}(\tilde{y})H_{m_x}(x)
 H_{m_y}(\tilde{y})V(x,\tilde{y}/\sqrt{\omega}) dxd\tilde{y}\Big) = \\
A_nA_m\Big( \sqrt{\pi}2^{n_y}n_y! \delta_{n_ym_y}I_P(n_x,m_x) 
+ \omega\sqrt{\pi}2^{n_x}n_x! \delta_{n_xm_x}I_P(n_y,m_y) 
+ \sum_{i+j\leq k} \alpha_{ij} \omega^{-j/2} I(n_x, m_x, i)I(n_y,m_y,j) \Big),
\end{multline}
where $I_P(n,m)$ is:
\begin{equation}
I_P(n,m) = - \frac{1}{2} I(n,m,2) + 1/2 I(n,m,0) + 2mI(n,m-1,1) - 2m(m-1)I(n, m-2, 0),
\end{equation}
being
\begin{equation}
I(n,m,r) = \int_{-\infty}^\infty x^r e^{-x^2} H_n(x) H_m(x) dx.
\end{equation}

\bibliography{mybibfile}

\end{document}